\documentclass{aa}  
\usepackage{graphicx}
\usepackage{txfonts}
\usepackage{footnote}
\usepackage[colorlinks=true,linkcolor=blue,citecolor=blue,filecolor=blue,urlcolor=blue]{hyperref}
\usepackage{subcaption} 
\usepackage{amssymb} 
\usepackage{ulem} 
\usepackage[dvipsnames]{xcolor} 
\usepackage{siunitx}
\usepackage{ gensymb } 
\usepackage{lscape} 
\usepackage[flushleft]{threeparttable}

\newcommand{\kms}{\mbox{km~s$^{-1}$}}

\usepackage{soul} 
\newcommand{\orcid}[1]{\href{https://orcid.org/#1}{\textcolor[HTML]{A6CE39}{\aiOrcid}}}
\usepackage{multirow}
\usepackage{multicol}
\usepackage{url}

\usepackage{fancyhdr}
\usepackage{natbib}
\bibpunct{(}{)}{;}{a}{}{,}

\begin{document} 
  \title{Discovery of optical and infrared accretion disc wind signatures in the black hole candidate MAXI~J1348--630}
  \titlerunning{Optical and infrared wind in MAXI~J1348--630}
  
    \author{G. Panizo-Espinar \inst{1}\textsuperscript{,}\inst{2}\fnmsep\thanks{E-mail: guayente.panizo@gmail.es}
       \and M. Armas Padilla\inst{1}\textsuperscript{,}\inst{2}
       \and T. Muñoz-Darias \inst{1}\textsuperscript{,}\inst{2} 
       \and K. I. I. Koljonen \inst{3}\textsuperscript{,}\inst{4}\textsuperscript{,}\inst{5}
       \and V. A. Cúneo \inst{1}\textsuperscript{,}\inst{2} 
       \and J. Sánchez-Sierras \inst{1}\textsuperscript{,}\inst{2}
       \and D. Mata Sánchez \inst{1}\textsuperscript{,}\inst{2}
       \and J. Casares \inst{1}\textsuperscript{,}\inst{2}         
       \and J. Corral-Santana \inst{6}
       \and R. P. Fender  \inst{7}\textsuperscript{,}\inst{8} 
       \and F. Jiménez-Ibarra \inst{9}
       \and G. Ponti \inst{10}\textsuperscript{,}\inst{11}
       \and D. Steeghs \inst{12}
       \and M.A.P.~Torres \inst{1}\textsuperscript{,}\inst{2}
               }
\authorrunning{G. Panizo-Espinar} 
   \institute{Instituto de Astrofísica de Canarias (IAC), Vía Láctea, La Laguna, E-38205, Santa Cruz de Tenerife, Spain
    \and 
    Departamento de Astrofísica, Universidad de La Laguna, E-38206 Santa Cruz de Tenerife, Spain 
    \and 
    Finnish Centre for Astronomy with ESO (FINCA), University of Turku, Väisäläntie 20, 21500 Piikkiö, Finland Aalto 
    \and 
    University Metsähovi Radio Observatory, P.O. Box 11000, FI-00076 Aalto, Finland
    \and     
    Institutt for Fysikk, Norwegian University of Science and Technology, Trondheim, Norway.
    \and
    European Southern Observatory, Alonso de Cordova 3107, Vitacura, Casilla 19001, Santiago de Chile, Chile
    \and
    Department of Physics, University of Oxford, Denys Wilkinson Building, Keble Road, Oxford OX1 3RH, UK
    \and
    Department of Astronomy, University of Cape Town, Private Bag X3, Rondebosch 7701, South Africa
    \and
    Australian Astronomical Optics, Macquarie University, 105 Delhi Rd, North Ryde, NSW 2113, Australia
    \and
    Osservatorio Astronomico di Brera, Via E. Bianchi 46, I-23807 Merate (LC), Italy
    \and
    Max-Planck-Institut fur Extraterrestrische Physik, Giessenbachstrasse, D-85748, Garching, Germany
    \and 
    Department of Physics, University of Warwick, Coventry CV4 7AL, UK
    }

  \abstract
  {MAXI~J1348--630 is a low mass X-ray binary discovered in 2019 during a bright outburst. During this event, the system sampled both hard and soft states following the standard evolution. We present multi-epoch optical and near-infrared spectroscopy obtained with X-shooter at the Very Large Telescope. Our dataset includes spectra taken during the brightest phases of the outburst as well as the decay towards quiescence. We study the evolution of the main emission lines, paying special attention to the presence of features commonly associated with accretion disc winds, such as blue-shifted absorptions, broad emission line wings and flat-top profiles. 
  We find broad emission line wings in H$\alpha$ during the hard-to-soft transition and blue-shifted absorption troughs at $\sim -500$ \kms \
  in H$\beta$, \ion{He}{i}--5876, H$\alpha$ and Pa$\beta$ during the bright soft-intermediate state. In addition, flat-top profiles are seen throughout the outburst. We interpret these observables as signatures of a cold (i.e. optical to infrared) accretion disc wind present in the system. We discuss the properties of the wind and compare them with those seen in other X-ray transients.  In particular, the wind velocity that we observe is low when compared to those of other systems, which might be a direct consequence of the relatively low binary inclination, as suggested by several observables. This study strengthen the hypothesis that cold winds are a common feature in low mass X-ray binaries and that they can also be detected in low inclination objects via high-quality optical and infrared spectroscopy.  
  } 
  

   \maketitle
    \section{Introduction}\label{cap.introduccion}
    Low mass X-ray binaries (LMXBs) are binary systems comprising a stellar-mass black hole (BH) or a neutron star (NS), that is accreting gas from a low mass donor star ($\lesssim$ 1 M$_\sun$).  The infalling material forms an accretion disc around the compact object (\citealt{ShakuraSunyaev1973}), whose internal regions reach temperatures high enough  (10$^{6}$--10$^{7}$ K) to emit in X-rays.
    The vast majority of LMXBs are transients. They spend most of their lives in a quiescent, low luminosity state, showing occasional accretion episodes when their luminosity increases by several orders of magnitude: the so-called outbursts.
    
    During outbursts, LMXBs usually evolve following a common X-ray hysteresis pattern (e.g. \citealt{Munoz-Darias2014}) that is related to their accretion properties. At the beginning of the outburst, the X-ray spectrum is dominated by a hard, power-law component thought to be produced by inverse-Compton processes in a corona of hot electrons (e.g. \citealt{Gilfanov2010}). This state is known as the hard state. As the outburst continues the system enters the soft state, when the spectrum becomes dominated by a soft, thermal component arising in the accretion disc. LMXBs can be also found in intermediate states, when they are transitioning between the above two main states \citep[see e.g. ][]{McClintock2006, Done2007, Belloni2011}. 
    
    Outbursts are also characterized by the presence of outflows, with  observational evidence indicating that they are directly correlated with the accretion state (for a review, see e.g. \citealt{Fender2016}). These outflows are usually observed in two flavours: jets and winds. Radio jets are ubiquitously observed in the hard (compact jet) and intermediate (discrete ejections) states, but not during the soft state \citep[e.g.][]{Gallo2003, Fender2004,Russell2011}. Conversely, hot X-ray winds are mainly detected during the soft state as blue-shifted X-ray absorptions in spectral transitions of highly ionized material, such as Fe \textsc{xxv} and Fe \textsc{xxvi} \citep{Ueda1998, Neilsen2009, Ponti2012, Ponti2014, DiazTrigo2016}. In addition, more recent studies have also shown the presence of colder accretion disc winds, which are mainly observed in optical emission lines and can be simultaneous with the jet (e.g. \citealt{Munoz-Darias2016}). These are revealed by some specific signatures: blue-shifted absorptions (P-Cygni profiles or absorption troughs), broad emission line wings, flat-top line profiles and line asymmetries. To date, they have been seen in a handful of BH (e.g. \citealt{Munoz-Darias2016, Munoz-Darias2018, Munoz-Darias2019, Cuneo2020b, MataSanchez2022a}) and NS transients (e.g. \citealt{Munoz-Darias2020}; see also \citealt{CastroSegura2022} for a detection in the ultraviolet). Furthermore, near-infrared studies show that these lower ionisation winds (a.k.a. cold winds) are likely active during most part of the outburst  \citep{Sanchez-Sierras2020}. Interestingly, both X-ray and optical/infrared (OIR, hereafter) wind detections occurred in systems with relatively high orbital inclinations, which is commonly interpreted as the wind having an equatorial geometry (e.g. \citealt{Ponti2012}; but see \citealt{Higginbottom2019}).

    MAXI~J1348--630 (J1348 hereafter) is a new LMXB, discovered during an outburst episode on 2019 January 26  \citep{Yatabe2019} by The Monitor of All-sky X-ray Image (\textit{MAXI}) nova alert system \citep{Matsuoka2009}. The peak X-ray flux was measured on February 10 ($\sim1.0 \times 10^{-7} \mathrm{erg~cm}^{-2}~\mathrm{s}^{-1}$, \citealt{Lamer2021, Carotenuto2021}), while the highest reported optical flux (\textit{g'} = 16.1) corresponds to February 5 \citep{Baglio2019}. A hard-to-soft transition began on February 3 \citep{Nakahira2019}.  X-ray studies using the Neutron Star Interior Composition Explorer (\textit{NICER}) revealed type-B quasi-periodic oscillations between February 8 and 26  (\citealt{Belloni2020, Zhang2021}, see also \citealt{Zhang2022}), associated with the soft-intermediate state (SIMS). The main outburst lasted $\sim$3 months and was followed by a few hard state-only re-brightenings (\citealt{Russell2019,Pirbhoy2020,Negoro2020,Baglio2020,Zhang2020, Corral-Santana2016}; see also \citealt{Cuneo2020}).
    Based on different spectral and timing properties, J1348 has been suggested to harbour a BH \citep{Sanna2019,Belloni2020, Zhang2020}, being located at $\sim$2--3~kpc \citep{Chauhan2021, Russell2019c, Lamer2021}.
	
	In this paper we present multi-epoch spectroscopy of J1348 obtained with X-shooter at the Very Large Telescope (VLT), covering ultraviolet, optical and near-infrared wavelengths. We study the presence/absence of the most typical LMXB emission lines, as well as the evolution of their main properties. 

    \begin{figure}[ht!]
    \centering
    \begin{subfigure}{\columnwidth}
        \includegraphics[width=9truecm]{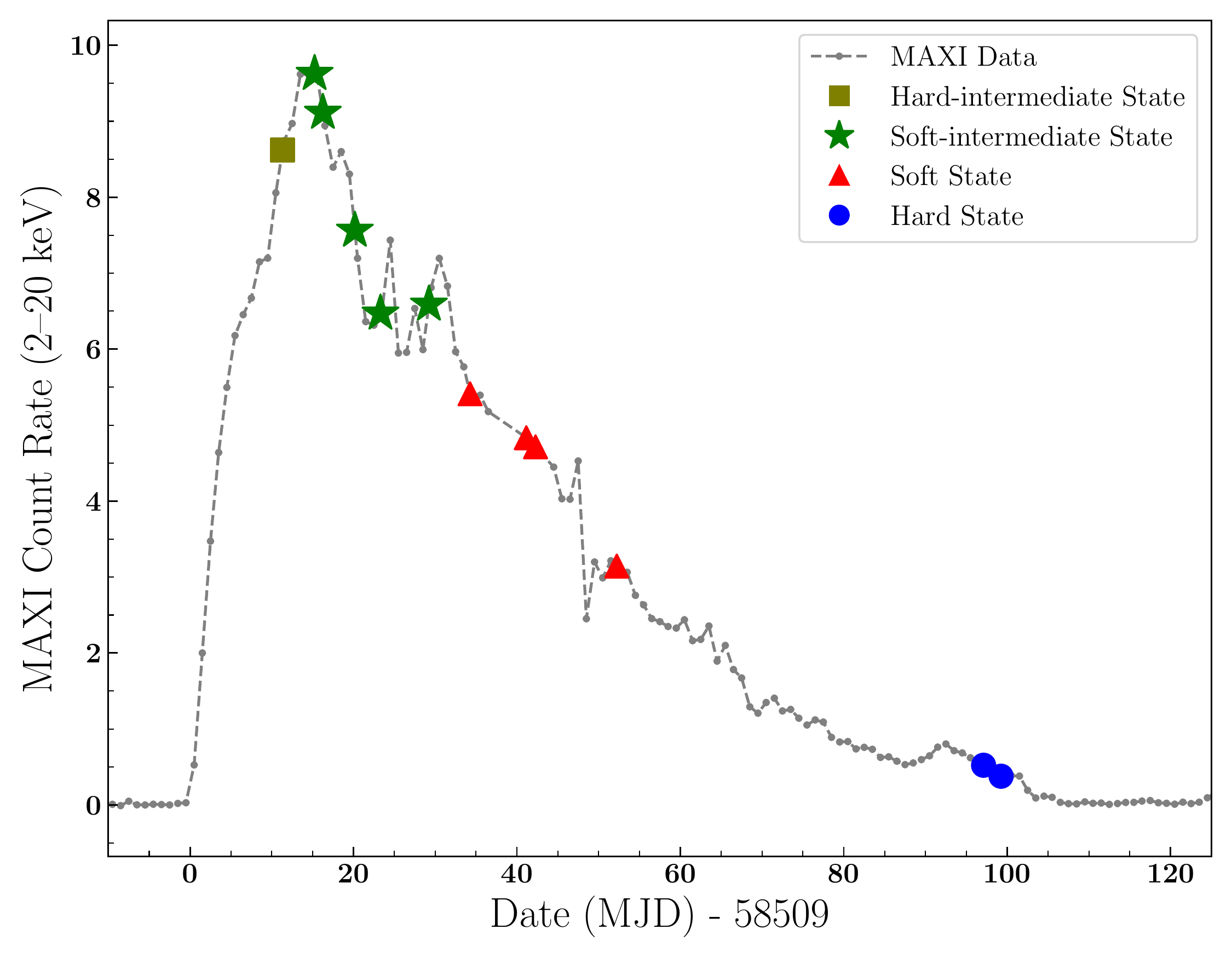}
    \end{subfigure}
    \begin{subfigure}{\columnwidth}
        \includegraphics[width=9truecm]{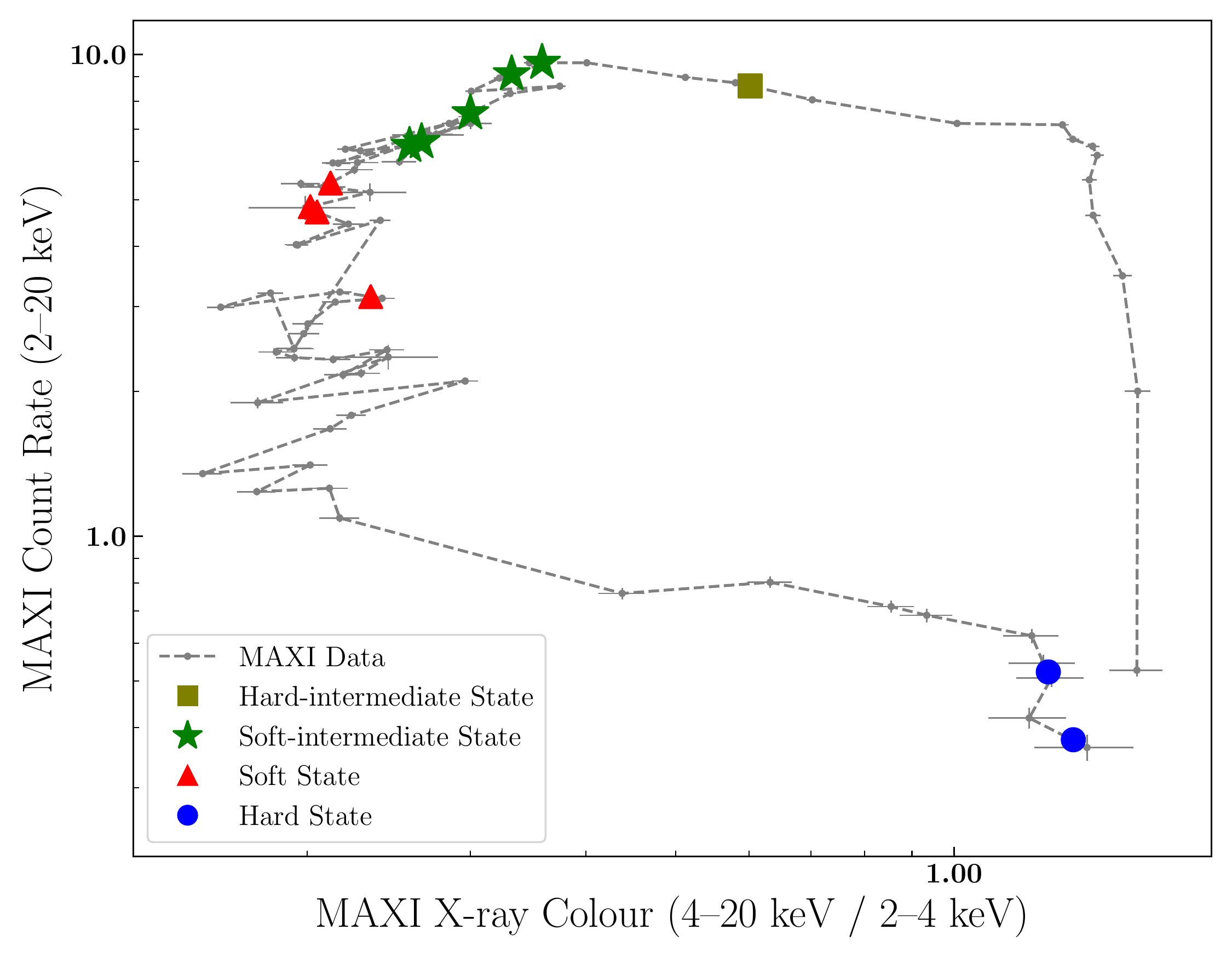}
    \end{subfigure}
    \caption{Light curve (upper panel) and HID (lower panel) of J1348 during its 2019 outburst (\textit{MAXI} data between MJD 58509 and 58634). The small, grey dots represent X-ray data, while the colored symbols mark the interpolated position for each X-shooter epoch. For the HID we only considered data with count rates exceeding 0.025 photons cm$^{-2}$ s$^{-1}$ in each band. The X-ray colour is defined as the ratio between the hard (4--20 keV) and the soft (2--4 keV) count rates. The source displays the classic hysteresis loop in the anticlockwise direction.}
    \label{fig1_light_HID}
    \end{figure}

	\begin{table*}[t!]
    	\centering
		\caption{Summary of the observing campaign}
		\begin{threeparttable}
		\begin{tabular}{c c c c c c c}
			\hline
			Epoch & 2019 Date (MJD) & X-ray State  & \textit{r}-band (\textit{i}-band) \tablefootmark{a} & \multicolumn{3}{c}{Total exposure time (s) / Slit width (")} \\
			 & & &  (AB magnitude) & UVB & VIS & NIR  \\
			\hline
			\hline
			1 & 06 Feb (58520) 	& Hard-intermediate & 15.4 & 2200 / 1.3 & 2000 / 1.2 & 2200 / 1.2  \\ 
			2 & 10 Feb (58524)  & Soft-intermediate & 15.4 & 4524 / 1.3, 1.0& 4116 / 1.2, 0.9 & 3360 / 1.2, 0.9   \\
			3 & 11 Feb (58525)  & Soft-intermediate & 15.5 & 2324 / 1.0 & 2216 / 0.9 & 1160 / 0.9 \\
			4 & 15 Feb (58529)  & Soft-intermediate & 15.6 & 2324 / 1.0 & 2216 / 0.9 & 1160 / 0.9  \\
			5 & 18 Feb (58532)  & Soft-intermediate & 15.6 & 3924 / 1.9, 1.0 & 3416 / 1.2, 0.9 & 2760 / 1.2, 0.9  \\
			6 & 24 Feb (58538)  & Soft-intermediate & -    & 1600 / 1.3 & 1200 / 1.2 & 1600 / 1.2 \\
			7 & 06 Mar (58543) 	& Soft              & -    & 2220 / 1.3 & 2000 / 1.2 & 2200 / 1.2  \\ 
			8 & 08 Mar (58550) 	& Soft              & 16.2 (15.1)& 2220 / 1.3 & 2000 / 1.2 & 2200 / 1.2  \\
			9 & 09 Mar (58551) 	& Soft              & 15.9 & 2324 / 1.0 & 2216 / 0.9 & 1160 / 0.9 \\
			10 & 19 Mar (58561) & Soft              & -    & 2200 / 1.3 & 2000 / 1.2 & 2200 / 1.2  \\
			11 & 03 May (58606) & Hard              & -    & 1920 / 1.3 & 2036 / 1.2 & 1920 / 1.2 \\
			12 & 05 May (58608) & Hard              &(16.9)& 1920 / 1.3 & 2036 / 1.2 & 1920 / 1.2  \\ 
			\hline
		\end{tabular}
        \end{threeparttable}
    \tablefoot{
    \tablefoottext{a}{The estimated uncertainty is $\sim$0.07 and $\sim$0.03 in \textit{r}-band and  \textit{i}-band, respectively.} 
    }
    \label{Table_data}
	\end{table*}

\section{Observations and data reduction}\label{cap.observations}
A total of 12 epochs of spectroscopy were obtained between February and May 2019 with X-shooter \citep{Vernet2011} at the VLT-UT2 in Cerro Paranal, Chile. We obtained photometric magnitudes for the epochs with available acquisition images, presented in the observing log (Table \ref{Table_data}). Each spectral observation included 4--16 exposures in nodding configuration with total exposure times of 1100--4500 s and covering from 3000 to 25000~\AA between the three arms of the instrument. 
We used slit widths of 1.0"--1.3", 0.9"--1.2" and 0.9"--1.2" 
for the ultraviolet (UVB), visible (VIS) and  infrared arms (NIR),  producing velocity resolutions of  $\sim$55--73~\kms, $\sim$34--46~\kms \ and $\sim$53--70~\kms, respectively. Airmasses ranged from 1.28 to 1.95. 

   \begin{figure*}[t!]
   \centering
   \includegraphics[width=18truecm]{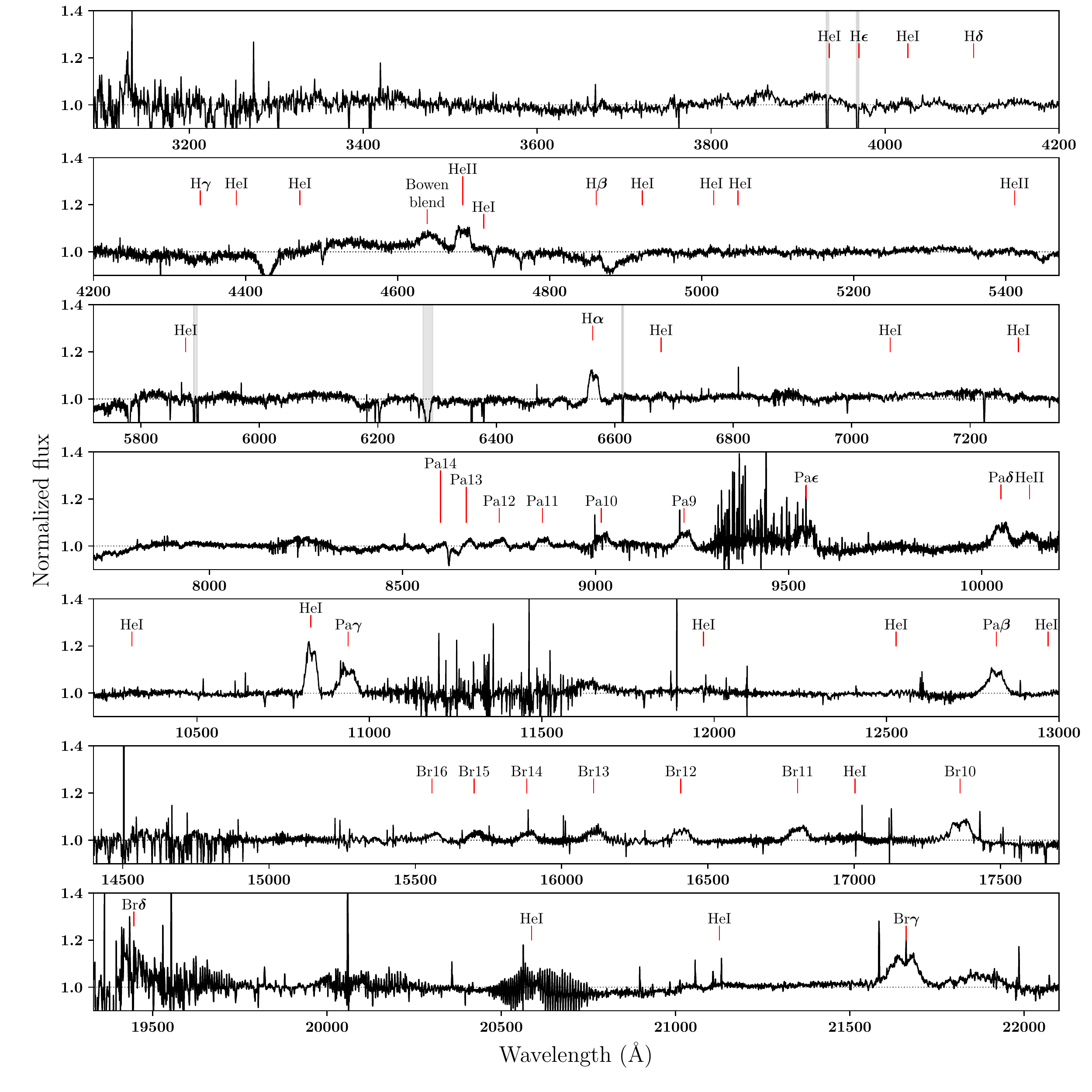}
    \caption{normalised OIR spectrum during epoch \#5. Relevant emission lines are labelled and prominent DIBs are shaded in grey.} 
	\label{fig_spectra_example}
    \end{figure*}

The spectra were reduced using the X-shooter ESO Pipeline v3.3.5. Flux calibration standard stars in epochs \#3--5 and \#8--9 were not bright enough in the infrared. Therefore, those spectra were calibrated with the standard star from epoch \#10. Atmospheric telluric absorptions in optical and infrared ranges were corrected using \textsc{molecfit} (\citealt{Smete2015}, \citealt{Kausch2015}), and the continuum adjacent to each spectral line of interest was individually normalised for a detailed study.

\section{Analysis and results}\label{cap.results}
J1348 was first detected on 2019 January 26 (MJD 58509) and it was X-ray active for $\sim$100 days. The upper panel in Fig.~\ref{fig1_light_HID} shows its X-ray light curve, which follows the usual fast rise ($\sim$10 days) and a slower decay towards quiescence. The lower panel in Fig.~\ref{fig1_light_HID} shows the hardness-intensity diagram (HID; \citealt{Homan2001}) of daily averaged count rates from \textit{MAXI}, where the X-ray colour is defined as the ratio between hard (4$-$20 keV) and soft (2$-$4 keV) count rates. The source displays the characteristic hysteresis pattern commonly observed in LMXBs during outbursts (\citealt{Munoz-Darias2014}). According to this diagram, we consider epoch \#1 to be in the hard-to-soft transition (most likely hard-intermediate state), \#2--6 in the soft-intermediate (type-B quasi-periodic oscillations reported in \citealt{Belloni2020,Zhang2020}), \#7--10 in the soft state and \#11--12 in the low-luminosity hard state, when the source was returning to quiescence. A summary of the observations and associated X-ray states is shown in Table \ref{Table_data}. 
   \begin{figure*}[ht!]
   \centering
   \includegraphics[width=\textwidth]{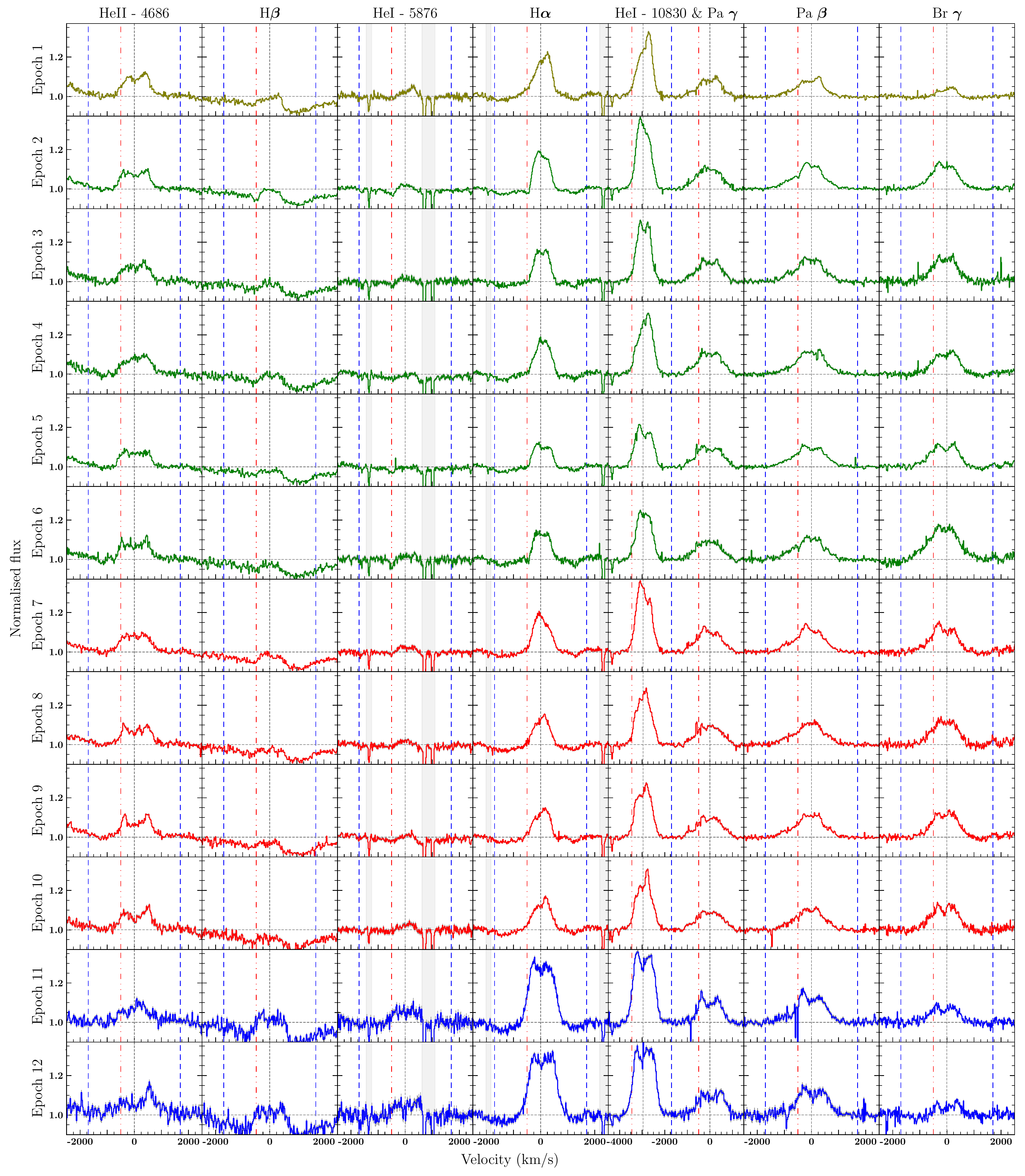}
   \caption{Evolution of the most relevant emission lines. Colour code is the same than in Fig. \ref{fig1_light_HID}. The blue dashed and red dashed-dotted vertical lines mark $\pm$1700~\kms \ and $-500$~\kms, velocities of the blue wing and the absorption trough of Pa$\beta$, respectively (in epochs \#2 and \#6). Grey bands indicate contaminated regions by DIBs or tellurics. The emission at $-$2500 \kms \ in \ion{He}{ii}--4686 column is produced by the Bowen Blend.} 
    \label{fig_grid}
    \end{figure*}  
   \begin{figure*}
    \centering
   \includegraphics[width=\textwidth]{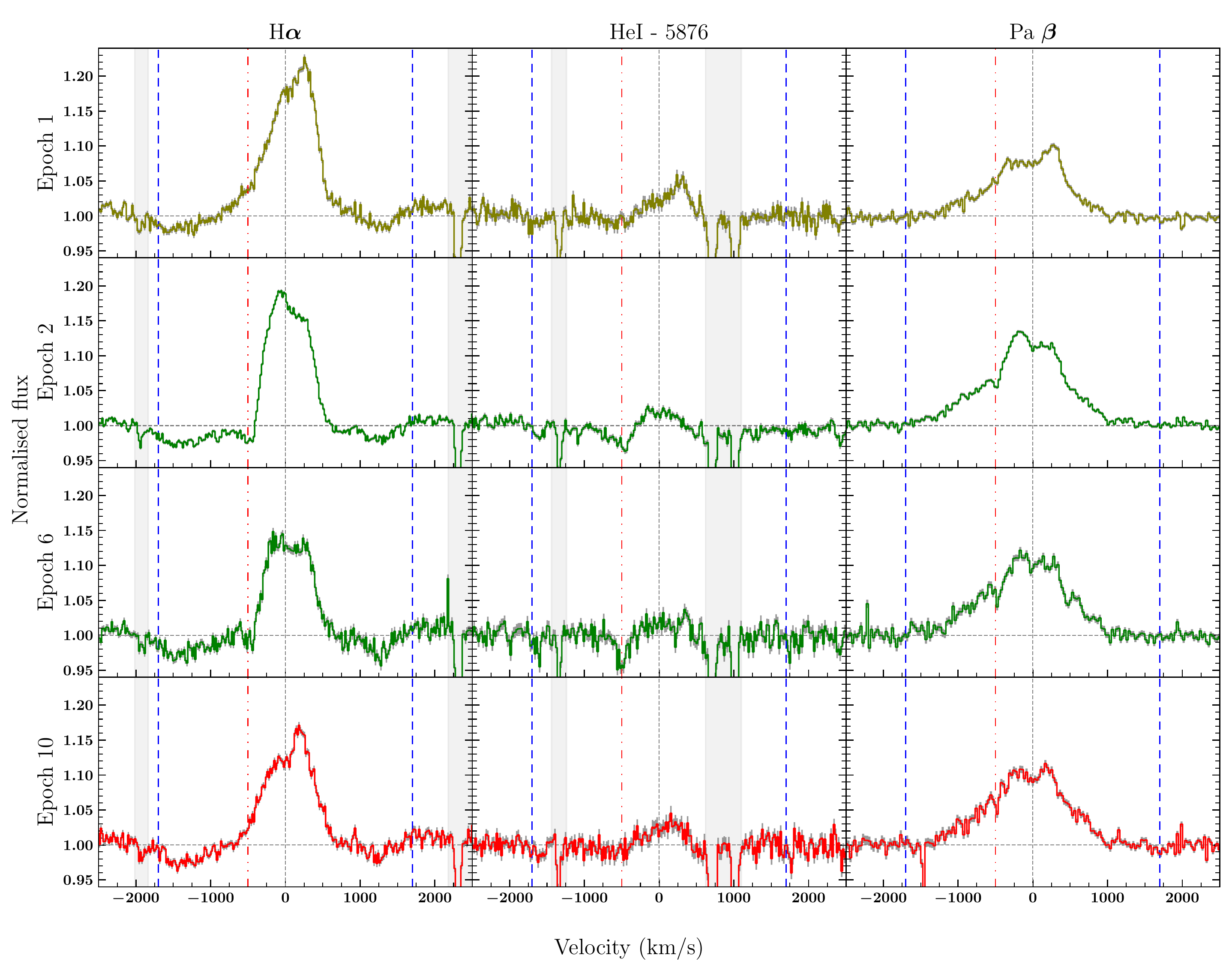}
    \caption{H$\alpha$, \ion{He}{i}--5876 and Pa$\beta$ profiles in epochs \#1, 2, 6 and 10 (rebinned by a factor of two). Vertical lines and colours are the same than in previous figures.}
    \label{fig_compare_ep2_3_heI5875_Ha_paB}
    \end{figure*}
   \begin{figure*}[ht!]
   \centering
   \includegraphics[width=\textwidth]{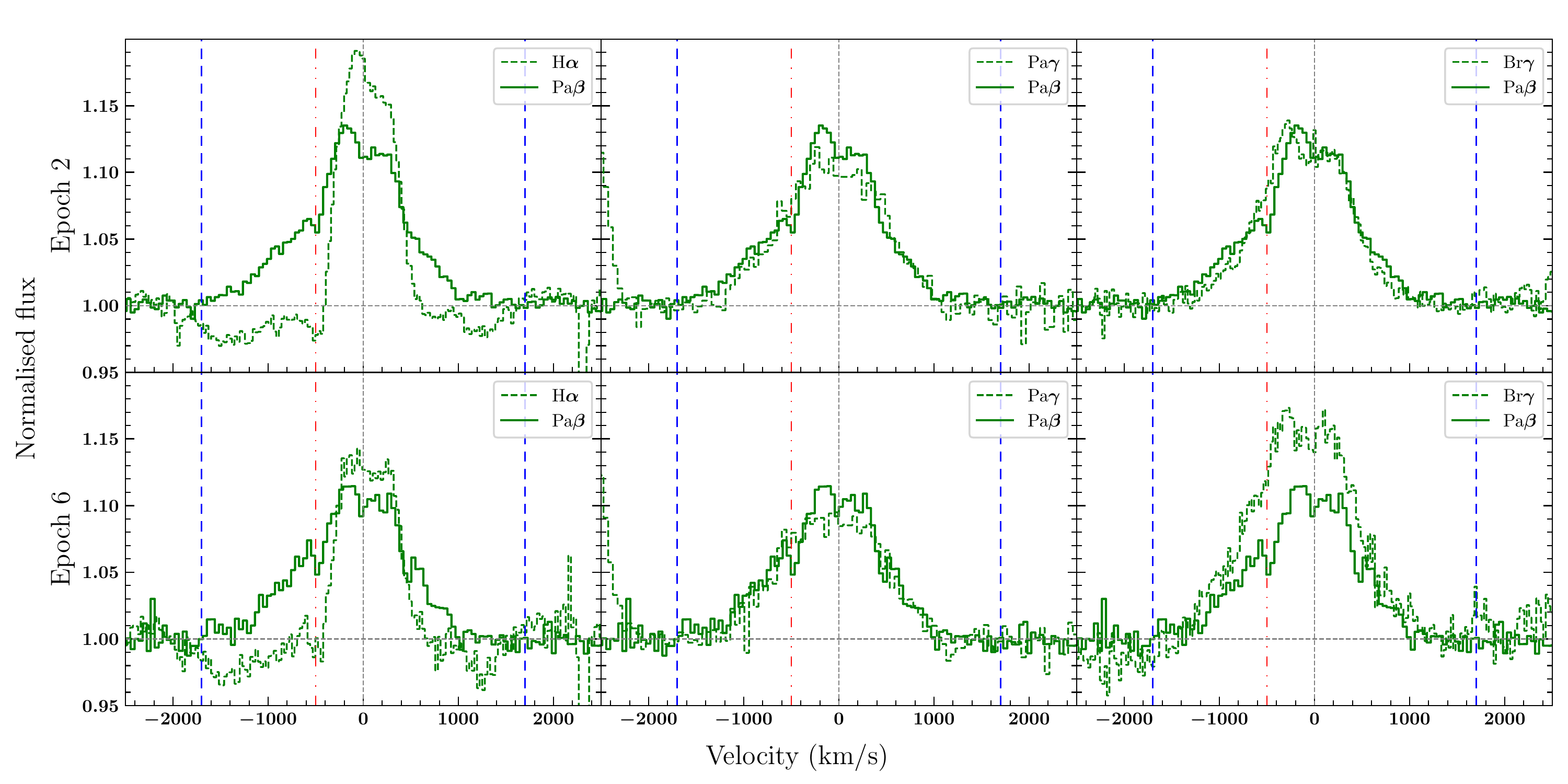}
    \caption{Pa$\beta$ profiles (solid line) compared to H$\alpha$, Pa$\gamma$ and Br$\gamma$ (dashed line) in epochs \#2 (upper panel) and \#6 (lower panel). Spectral lines are rebinned by a factor of 3 for clarity. Vertical lines and colours are the same than in previous figures.}
    \label{fig_compare_paB_paG_bG}
    \end{figure*}  

\subsection{Spectral evolution}\label{subsect.specEvol}
The OIR spectrum of J1348 varied as the outburst evolved, with the most prominent emission lines observed during the last epochs (i.e. hard states; \#11--12). 
We observe several hydrogen emission lines in the broad spectral range covered, from the Balmer, Paschen, and Brackett series (as an example, see epoch \#5 spectra in Fig. \ref{fig_spectra_example}). Helium lines such as \ion{He}{ii} at 4686~\mbox{\normalfont\r{A}} (\ion{He}{ii}--4686),  \ion{He}{i}--5876, \ion{He}{i}--6678 and \ion{He}{i}--10830 are also visible. Figure \ref{fig_grid} shows the evolution of the main emission lines across the observing campaign. All of them were individually normalised by fitting a first-order polynomial to their local continuum with the emission line masked (and the underlying absorption also masked in Balmer lines). 

H$\alpha$ shows intense, double-peaked profiles in many epochs. The peak-to-peak separation increases as the outburst evolves, from 310~$\pm$~150~\kms \  in epoch \#1 to a maximum of 600~$\pm$~70~\kms \ in the last epoch. This behaviour, revealed by two-Gaussian models fitted to the line (with all parameters free to vary), has been observed in other systems and interpreted as an indication of disc shrinking towards the end of the outburst (e.g. \citealt{Casares2019, MataSanchez2015, Torres2015, CorralSantana2013}). This line shows an asymmetric profile and broad wings reaching $\sim$~$\pm$1500 \kms \ in epoch \#1 (see Fig.~\ref{fig_grid} and \ref{fig_compare_ep2_3_heI5875_Ha_paB}), while a flat-top profile is visible in epoch \#6. An underlying broad absorption component (from $\sim$--2000 to $\sim$1700 \kms) is also present during most epochs, a feature that is relatively commonly observed in LMXBs in outburst \citep{Soria2000, Casares1995, Yao2021}. This is also present in H$\beta$ and H$\gamma$ (see e.g. in Fig. \ref{fig_spectra_example}).

H$\beta$ shows emission profiles similar to H$\alpha$, although much less intense. This line also shows broad absorptions in every epoch, particularly clear in \#5--12 ($\sim$10\% below the continuum reaching $\sim$ $\pm$3000 \kms). This specially prominent and red-shifted absorption has been observed in other transients in outburst \citep[e.g.][]{Cuneo2020b} and is most likely a combination of the aforementioned Balmer broad absorption component and an additional contribution from a broad diffuse interstellar band (DIB) at $\sim$4882 \mbox{\normalfont\r{A}} with FWHM $\sim$20 \mbox{\normalfont\r{A}}  (e.g. \citealt{Kaur2012, Jenniskens1994,Buxton2003}). 

Pa$\beta$ profiles are qualitatively similar but broader and less intense than H$\alpha$. In addition, several lines from the Paschen series show flat-top-like profiles in epochs \#4, \#6 and \#9. Similar boxy line profiles are also observed in high-order Brackett lines during epochs \#2 and \#4 (e.g. Br-10, 11, 12, similar to those seen in other transitions; see Fig. \ref{fig_grid}). Br$\gamma$ shows profiles similar to those of Pa$\beta$ (Fig. \ref{fig_grid}) except for the less prominent ones in harder epochs (\#1, \#11--12), as observed in other transients \citep{Sanchez-Sierras2020}.

Finally, \ion{He}{i}--10830 is generally the most intense He line. It shows line profiles very similar to those of H$\alpha$, but with more prominent double-peaks. Its peak-to-peak separation also increases as the outburst evolves, from 400 $\pm$ 60 \kms \ to a maximum of 590 $\pm$ 90 \kms \ in epoch \#12. Other typical He lines, such as \ion{He}{i}--5876  and \ion{He}{ii}--4686 are also visible. The latter, which is expected to trace hotter (inner) gas, shows double-peaked (characteristic of gas rotating in an accretion disc, e.g. \citealt{Smak1969}) or boxy profiles depending on the epoch (see Fig. \ref{fig_grid}).

\subsection{Outflow features: Blue-shifted absorption troughs and broad wings}\label{subsect.AbsTro}
As mentioned above, during some stages of the outburst several spectral lines display features that could be interpreted as signatures of accretion disc winds. These include blue-shifted absorptions, broad emission line wings, and line profiles with asymmetries or flat-tops. Fig.~\ref{fig_grid} and \ref{fig_compare_ep2_3_heI5875_Ha_paB} show the presence of a blue-shifted absorption trough centred at $-$500~\kms\ in epochs \#2 and \#6 in H$\beta$, \ion{He}{i}--5876, H$\alpha$ and Pa$\beta$ (red dashed-dotted line). In the first three cases this  feature resembles the blue-shifted absorption of a P-Cygni profile. 
However, due to the broader profile of Pa$\beta$, the absorption trough is observed as a dent on its blue wing. Figure~\ref{fig_compare_paB_paG_bG} shows that this feature is not visible in Pa$\gamma$ nor Br$\gamma$. It is worth noting that the blue-shifted absorption becomes visible again in the soft state (epoch \#10), but in this case only in Pa$\beta$ (see Fig.~\ref{fig_compare_ep2_3_heI5875_Ha_paB}). Interestingly, the profile of Pa$\beta$ in  epochs \#2 and \#6 also suggests the presence of an underlying broader emission component, with its blue wing reaching $\sim -$1700~\kms\ (blue vertical dashed line  in Fig.~\ref{fig_grid}, \ref{fig_compare_ep2_3_heI5875_Ha_paB} and \ref{fig_compare_paB_paG_bG}) and the red one $\sim$ +1000~\kms. This velocity is similar to that of the aforementioned broad absorption components best seen in H$\alpha$ in epoch \#1 ($\sim$ $\pm$1700 \kms, see Fig.~\ref{fig_compare_paB_paG_bG} ). 

   \begin{figure}[ht!]
   \centering
   \includegraphics[width=\columnwidth]{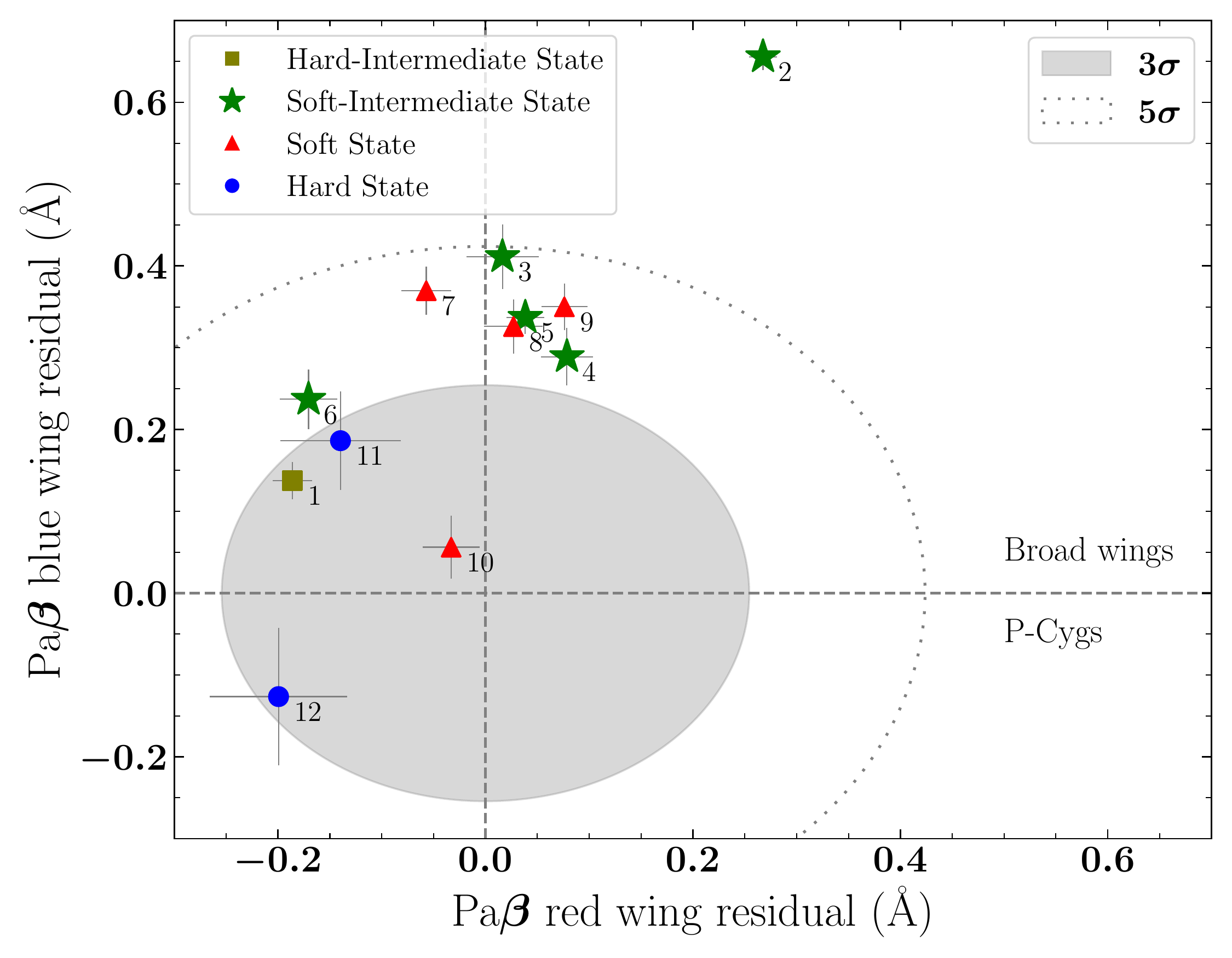}
    \caption{Pa$\beta$ Excesses Diagnostic Diagram. Symbol and colour codes are the same as in Fig. \ref{fig1_light_HID}. The grey region and dotted line indicate the 3- and 5-$\sigma$ contours, respectively. Epochs showing wind-related signatures are expected to lie on the right quadrants.}
    \label{fig_EDD}
    \end{figure}  

\subsubsection{The Excesses Diagnostic Diagram}\label{EDD}
Some of the spectral features revealed by visual inspection suggest the presence of an outflow in the system (see above). In a second step, we performed a systematical inspection using the Excesses Diagnostic Diagram. This tool was introduced by \citet{MataSanchez2018}, and is based on studying the deviation of the spectral line wings from a pure Gaussian profile. The coordinates of each epoch in the diagram are the so-called red and blue excesses (EW$_{r+}$ and EW$_{b-}$), defined as the equivalent width (EW) of the line wings after subtracting the Gaussian model (see \citealt{Munoz-Darias2019, Panizo-Espinar2021} for details). Thus, an epoch with a P-Cygni profile (i.e. red emission and blue absorption, thus EW$_{r+}$\textgreater0 and EW$_{b-}$\textless0) would lie on the bottom-right quadrant, while broad wings would make it fall in the top-right one (i.e. red and blue emission, EW$_{r+}$\textgreater0 and EW$_{b-}$\textgreater0).

Although this method is generally applied to H$\alpha$, in this case the results are non-conclusive due to the variable contamination by the underlying broad absorption component. We tried a multi-Gaussian fitting to model the broad absorption but it did not provide meaningful results. Pa$\beta$, on the other hand, shows significant (>5$\sigma$), positive red and blue wing residuals during epoch \#2, measured in the ($\pm$)700--1500 \kms\ velocity bands (Fig. \ref{fig_EDD}). This is consistent with the presence of broad wings, as  suggested by the visual inspection. No P-Cygni profiles are revealed by the diagram, which is an expected outcome given that the blue-shifted absorptions appear as small dents (with small equivalent width) on Pa$\beta$ (see Sect. \ref{subsect.AbsTro}).

\section{Discussion}\label{sect.discussion}
We have presented OIR spectroscopy of J1348 during its 2019 discovery outburst. 
We studied the evolution of the main spectral lines throughout the entire accretion episode, paying special attention to those particularly sensitive to outflows. We find several spectral features commonly associated with the presence of an accretion disc wind.

Firstly, a blue-shifted absorption trough is seen at $-$500~\kms \ during the bright soft-intermediate state (epoch \#2 and \#6) in H$\beta$, \ion{He}{i}--5876, H$\alpha$ and Pa$\beta$  (see Fig.~\ref{fig_compare_ep2_3_heI5875_Ha_paB} and ~\ref{fig_compare_paB_paG_bG}). The fact that this narrow feature is simultaneously observed at the same velocity in four different lines rules out an artifact (e.g. deficient sky subtraction) or a DIB origin. This absorption at $-$500~\kms\ is present again in Pa$\beta$ during the soft state (epoch \#10, see Fig.~\ref{fig_compare_ep2_3_heI5875_Ha_paB}).

Secondly, broad emission line wings are spotted in H$\alpha$  during epoch \#1 reaching $\sim$~$\pm$1500 \kms, and in Pa$\beta$ during epoch \#2 reaching $\sim$~-1700 \kms \ (see Fig. \ref{fig_compare_ep2_3_heI5875_Ha_paB} and Sect. \ref{EDD}). Profiles with broad wings can be interpreted as signatures of outflows (e.g. \citealt{Munoz-Darias2018, Panizo-Espinar2021}; see also \citealt{Prinja1996} for winds in massive stars).

Finally, several Balmer and helium lines, such as H$\alpha$, \ion{He}{i}--5876 and \ion{He}{i}--10830, are asymmetric with a negative skew (i.e., the emission peak appears red-shifted compared to the rest wavelength) in epoch \#1 (hard-intermediate state; Fig. \ref{fig_grid}). Similar profiles are commonly seen in systems showing conspicuous signatures of outflows (e.g. \citealt{Munoz-Darias2019}).

Additionally, flat-top profiles are observed in epoch \#2 in several hydrogen lines of the Paschen and Brackett series (Pa$\gamma$, Pa$\delta$, Pa-9, Pa-10, Pa-11, Br-10, Br-11), epoch \#6 in H$\alpha$, Paschen and helium lines (\ion{He}{i}--5876 and \ion{He}{i}--10830), and  epochs \#9--10 in Paschen and Brackett series (see Fig. \ref{fig_grid}), suggesting  the presence of the wind through the soft state. These profiles (albeit with a weak central absorption) are also present in the Balmer series and \ion{He}{i} lines during the final hard state epochs \#11--12 (Fig. \ref{fig_grid}). Similar features are commonly detected in novae and Wolf-Rayet stars \citep{Beals1931, VainuBappu1954, Williams2010} and generally understood as a signature of expanding envelopes (i.e. outflows; e.g. \citealt{Soria2000}).

Overall, at least four different wind-related spectral features are visible in J1348: blue-shifted absorption troughs, broad emission line wings, asymmetries and flat-top profiles, which appear simultaneously in several spectral lines. We interpret these features as evidence of an OIR accretion disc wind present during most of the brightest phase of the outburst. Unfortunately, we note that we were not able to monitor the system during the initial hard state, and in particular during the hard state peak, a stage in which other systems have shown the most conspicuous optical wind signatures (e.g. MAXI~J1820+070, \citealt{Munoz-Darias2019}; GRS~1716$-$249, \citealt{Cuneo2020}; MAXI~J1803$-$298, \citealt{MataSanchez2022a}). We also note that the possible presence of accretion disc winds in J1348 during previous activity of the system would support the hypothesis of them contributing to create a low density cavity in which the system seems to be embedded \citep{Carotenuto2021}, as suggested by other authors \citep{Carotenuto2022}.

\subsection{Comparison with other systems}
OIR accretion disc winds have been detected in a number of LMXBs during outbursts, as we summarize in Table \ref{Table_winds}. Most cases are BH transients with conspicuous wind features present during the hard state, although there are also examples of OIR winds found in NS systems \citep{Bandyopadhyay1999, Munoz-Darias2020}. In particular, most detections correspond to P-Cygni profiles observed in high inclination (\textgreater $60^\circ$) sources, with blue-edge velocities indicating relatively high velocity outflows (1200--3000~\kms). This prototypical wind signature is not particularly obvious in our data set. However, we find weak, blue-shifted absorption troughs in several lines in epochs \#2 and \#6. In particular, the blue-edge velocity 
of this absorption in \ion{He}{i}--5876 points to a terminal velocity of $\sim$900~\kms, smaller than the 
OIR winds velocities measured in other LMXBs (Table \ref{Table_winds}). There are also other (somewhat more tentative) wind features (e.g. broad emission line wings in \citealt{Rahoui2014, Soria2000, Panizo-Espinar2021}), which typically reach higher velocities than those inferred from the P-Cygni profiles \citep[see e.g.][]{Munoz-Darias2018, Munoz-Darias2019}. This is also the case for J1348, where the emission line wings of H$\alpha$ reach considerably higher values ($\sim$1500 \kms) than the narrow, blue-shifted absorptions (500--900 \kms). This might be explained by a dissimilar dependence of the various wind signatures to the orbital inclination, with blue-shifted absorptions being particularly sensitive to the presence of ouflowing material along the line-of-sight (see Sect. \ref{sect.discussion}).

Among the systems in Table \ref{Table_winds}, the comparison with MAXI~J1820+070 is particularly relevant. This BH transient showed a variety of wind signatures during its 2018 outburst, which also followed a standard evolution with hard and soft states. Specifically, Pa$\beta$ and Pa$\gamma$ showed absorption troughs with kinetic properties (wind velocity of 1200--2000 \kms) similar to those derived from P-Cygni profiles present in other spectral lines \citep{Sanchez-Sierras2020}. This is consistent with the behaviour of J1348, where the absorption features in Pa$\beta$ indicate the same velocity  than those found in H$\alpha$ and \ion{He}{i}--5876. 

\begin{table*}[ht!]
	\addtolength{\tabcolsep}{-3pt}
	\centering
	\caption{Characteristics of LMXBs with OIR winds and properties of the outflow}
    \begin{threeparttable}
	\begin{tabular}{c c c c c c c c }
		\multicolumn{8}{c}{(a) OIR wind detections} \\
        \hline
		 LMXB & Object  & \textit{i} ($\degree$) & P (h) & Wind features &  Vel. (\kms)\tablefootmark{a} & Spectral range  & References\\
        \hline
        \hline
        GX13+1 & NS & 60--80 & 589 & P-Cygni & $\sim$2400 & NIR & 1, 1b, 1c, 1d \\
        \hline
        \multirow{2}{*}{V404 Cyg} & \multirow{2}{*}{BH}  & \multirow{2}{*}{60--70}  & \multirow{2}{*}{155.3} & P-Cygni  & 1500--3000   & \multirow{2}{*}{Optical} &  \multirow{2}{*}{2, 2b, 2c, 2d, 2e}\\ 
          &    &      &      & Broad wings & $\sim$3000     &      & \\ 
		\hline
		\multirow{2}{*}{V4641 Sgr} & \multirow{2}{*}{BH} & \multirow{2}{*}{60--70} & \multirow{2}{*}{67.6} & P-Cygni  & 900--1600 & \multirow{2}{*}{Optical} &  \multirow{2}{*}{3, 3b, 3c, 3d, 3e}\\
		          &    &      &      & Broad wings & $\sim$3000     &      & \\ 
		\hline
		Swift J1357.2-0933  & BH  & $\gtrsim$80 & 2.8 & P-Cygni   & 1600--4000    & Optical  & 4, 4b, 4c\\ 
		\hline
		\multirow{3}{*}{MAXI~J1820+070} & \multirow{3}{*}{BH} & \multirow{3}{*}{67--81} & \multirow{3}{*}{16.4} & P-Cygni & 1200--1800    & Optical  &  \multirow{3}{*}{5, 5b, 5c, 5d}\\ 
		    &    & & & Broad wings & $\sim$1800     & Optical    & \\ 
		    &  &   &    & Blue-shifted abs.\tablefootmark{b}   & 1200--1800 & NIR    & \\ 
		\hline
		Swift J1858.6-0814  & NS  & Dipping/eclipsing   & 21.8 & P-Cygni & 1700--2400 & Optical & 6, 6b, 6c\\ 
		\hline
		\multirow{2}{*}{GRS1716-249}& \multirow{2}{*}{BH}  & \multirow{2}{*}{?} & \multirow{2}{*}{?} & P-Cygni & $\sim$2000& \multirow{2}{*}{Optical} & \multirow{2}{*}{7}\\
		         &   &   &   & Flat-top, asymmetries    & - &  &   \\ 
        
        \hline
		MAXI~J1803--298 & BH & Dipping & 7--8 & P-Cygni & $\sim$1250 & Optical & 8, 8b, 8c\\ 
		\hline
		\multirow{3}{*}{\textbf{MAXI~J1348--630}} & \multirow{3}{*}{\textbf{BH}} & \multirow{3}{*}{\textbf{ Mid-to-low} } & \multirow{3}{*}{\textbf{?}} & \textbf{Broad wings} & \textbf{1500--1700}   & \textbf{Optical, NIR} &  \multirow{3}{*}{\textbf{this work, 9}}\\ 
		  &  &  &  & \textbf{Blue-shifted abs.}  & \textbf{500--900} & \textbf{Optical, NIR}  &  \\ 
		  & &  &  & \textbf{Flat-top, asymmetries} & -  & \textbf{Optical, NIR}  & \\ 
		
        \hline
        \\
		\multicolumn{8}{c}{(b) Tentative OIR wind detections} \\
		\hline
		 LMXB & Object & \textit{i} ($\degree$) & P (h) & Wind features &  Vel.(\kms)\tablefootmark{a} & Spectral range  & References\\
        \hline
        \hline
		 Sco X-1  & NS & \textless40 & 18.9 & P-Cygni & $\sim$2600 & NIR  &   1, 10, 10b, 10c \\ 
		\hline
		GRO J1655--40 & BH & 67--70 & 62.9 & Flat-top, asymmetries & - & Optical & 11, 11b, 11c \\
		\hline
		GX339--4 & BH & 37--78 & 42.2 & Broad  wings & ? & NIR & 12, 12b \\
		\hline
		Aql X-1 & NS & 23--53 & 18.7 & Broad wings & 800 & Optical &  13, 13b, 13c\\ 
		\hline
    \end{tabular}
\end{threeparttable}

\tablefoot{
\tablefoottext{a}{Terminal velocity.}
\tablefoottext{b}{Absorption.}
}
\tablebib{           
(1) \citet{Bandyopadhyay1999}; 
(1b) \citet{Iaria2014}; 
(1c) \citet{DiazTrigo2012}; 
(1d) \citet{Homan2016}; 
(2) \citet{Munoz-Darias2016}; 
(2b) \citet{Munoz-Darias2017}; 
(2c) \citet{MataSanchez2018}; 
(2d) \citet{Casares1991}; 
(2e) \citet{Khargharia2010};  
(3) \citealt{Munoz-Darias2018}; 
(3b) \citet{Lindstrom2005};
(3c) \citet{Chaty2003};
(3d) \citet{Orosz2001}; 
(3e) \citet{MacDonald2014}; 
(4) \citet{JimenezIbarra2019b}; 
(4b) \citet{MataSanchez2015}; 
(4c) \citet{CorralSantana2013}; 
(5) \citet{Munoz-Darias2019}; 
(5b) \citet{Sanchez-Sierras2020}; 
(5c) \citet{Torres2019}; 
(5d) \citet{Torres2020};  
(6) \citet{Munoz-Darias2020}; 
(6b) \citet{Buisson2020b}; 
(6c) \citet{Buisson2021}; 
(7) \citet{Cuneo2020b}; 
(8) \citet{MataSanchez2022a}; 
(8b) \citet{Homan2021}; 
(8c) \citet{Xu2021}; 
(9) \citet{Carotenuto2022}; 
(10) \citet{Fomalont2001}; 
(10b) \citet{Gottlieb1975};  
(10c) \citet{MataSanchez2015b}; 
(11) \citet{Soria2000}; 
(11b) \citet{VanDerHooft1998}; 
(11c) \citet{Beer2002}; 
(12) \citet{Rahoui2014}; 
(12b) \citet{Heida2017}; 
(13) \citet{Panizo-Espinar2021}; 
(13b) \citet{MataSanchez2017}; 
(13c) \citet{Chevalier1991}.
}
\label{Table_winds}
\end{table*}

\subsection{On the state dependence of the wind features}\label{subsect.states}
The comparison of J1348 with other LMXBs allows us to revisit the state dependency of the wind-related OIR features. In J1348, the most compelling evidence for winds is arguably found during the SIMS, while the strongest optical winds are usually found at the hard state peak. A possible exception would be the first soft observation of MAXI~J1820+070 \citep{Munoz-Darias2019}, which was obtained right after the hard-to-soft transition (i.e. similar to observation \#2 in this paper).

Our observations did not cover the bright hard state. Nonetheless, additional OIR wind features were observed during the hard-intermediate (e.g. strong asymmetries in H$\alpha$ and \ion{He}{i}--10830) and the final (low-luminosity) hard state (flat-top profiles in the Balmer and \ion{He}{i} lines).  Wind-related features are only detected in the infrared lines (e.g. blue-shifted absorptions in Pa$\beta$ and flat-top profiles in the Paschen and Brackett series) during the soft state. This is consistent with the results (from MAXI~J1820+070) presented by \citealt{Sanchez-Sierras2020}. They reported on the detection of infrared wind-related features during most part of the outburst, whereas the optical ones were absent during the soft state. Thus, our results reinforce a scenario in which accretion disc winds are active during the whole outburst, although they are spotted at different wavelengths depending on the X-ray state. This dependence might be related to ionisation effects, making the infrared lines more sensitive to winds during the soft state (see \citealt{Sanchez-Sierras2020} and references therein).

\subsection{On the orbital inclination}\label{subsect.inclination}
J1348 does not show dips or eclipses, which suggests it is not a high inclination system. In fact, a mid-to-low inclination is supported by the orientation of the jet (\textit{i} = 29$\degree$ $\pm$ 3$\degree$, \citealt{Carotenuto2022}) assuming that there is not a strong misalignment between the jet and the orbital plane.
Although our analysis is not focused on this issue, the HID (Fig. \ref{fig1_light_HID}) supports this conclusion, since it shows a squared q-shape evolution (with an almost horizontal hard-to-soft transition), a suggested property of BH LMXBs with mid-to-low inclinations \citep{Munoz-Darias2013}.

In addition, some spectral lines, such as H$\alpha$ and \ion{He}{i}--10830 (Fig. \ref{fig_grid}), are double-peaked, which might be interpreted as a relatively high inclination feature. However, peak-to-peak separations for H$\alpha$ seem to be narrower 
(300--600 \kms, see Sect. \ref{cap.results}) than those measured in high inclination LMXBs (e.g. $\sim$1900~\kms \ in MAXI~J1659$-$152, \citealt{Torres2021}, $\sim$1200~\kms \ in XTE~J1118+480, \citealt{Torres2002} or 500$-$1100~\kms \ in MAXI~J1803$-$298, \citealt{MataSanchez2022a}). 

Overall, a mid-to-low inclination is favoured by all the available observables. This might explain the peculiar observational properties of the wind in this object. For instance, a predominantly equatorial wind would be detected with a lower (projected) velocity in a low inclination system. In addition, this would make more difficult to form blue-shifted absorptions that reach fluxes below the continuum level, since the contribution from the accretion disc emission lines is larger at low velocities (i.e. as the blue-shifted absorption moves closer to the core of the disc line). Therefore, unless the wind features are very strong, one might naturally expect to detect them as low velocity absorption troughs, qualitatively similar to those reported in this work.

\section{Summary and conclusions}\label{cap.conclusions}
We presented OIR spectroscopy of the BH candidate MAXI~J1348--630 obtained during its discovery outburst in 2019. We reported the detection of several spectral features that are commonly associated with the presence of outflows. In particular, we found:
\begin{itemize}
    
    \item Blue-shifted absorption troughs during  the soft-intermediate state (H$\beta$, \ion{He}{i}--5876, H$\alpha$ and Pa$\beta$) and the soft state (Pa$\beta$). These are centred at $-$500~\kms.
    
    \item Broad emission line wings (H$\alpha$ and Pa$\beta$) reaching velocities of up to $\pm$1700 \kms, as well as asymmetric lines with a negative skew and flat-top profiles. These are seen at different stages of the outburst.

\end{itemize}

We interpret these features as signatures of a cold (OIR) accretion disc wind that is present during, at least, the brightest phases of the accretion episode. The apparently lower velocity of the blue-shifted absorptions as compared to that seen in other LMXBs might be related to the mid-to-low orbital inclination of the system. Further spectroscopic observations of MAXI~J1348--630 are encouraged during both quiescence and forthcoming outbursts in order to determine its fundamental parameters and better characterise the properties of the wind.

\begin{acknowledgements}
    We are thankful to the anonymous referee for constructive comments that have improved this paper. DMS and MAP acknowledge support from the Consejer\'ia de Econom\'ia, Conocimiento y Empleo del Gobierno de Canarias and the European Regional Development Fund (ERDF) under grant with reference ProID2020010104 and ProID2021010132. TMD and MAPT acknowledge support via the Ram\'on y Cajal Fellowships RYC-2015-18148 and RYC-2015-17854, respectively. KIIK acknowledges funding from the European Research Council (ERC) under the European Union’s Horizon 2020 research and innovation programme (grant agreement No. 101002352) and from the Academy of Finland projects 320045 and 320085. GP acknowledges funding from the European Research Council (ERC) under the European Union’s Horizon 2020 research and innovation programme (grant agreement No 865637). This work has been supported in part by the Spanish Ministry of Science under grants AYA2017- 83216-P, PID2020-120323GB-I00 and EUR2021-122010. \textsc{Molly} software developed by Tom Marsh is gratefully acknowledged. Based on observations collected at the European Southern Observatory under ESO programmes  0102.D-0309(A) and 0102.D-0799(A). This research has made use of \textit{MAXI} data provided by RIKEN, JAXA and the \textit{MAXI} team. We acknowledge the use of public data from the \textit{Swift} data archive. 
\end{acknowledgements}	

%
%
\bibliographystyle{aa}
\bibliography{J1348.bbl}

\begin{thebibliography}{95}
\expandafter\ifx\csname natexlab\endcsname\relax\def\natexlab#1{#1}\fi

\bibitem[{Baglio {et~al.}(2019)Baglio, Russell, Bramich, \& Lewis}]{Baglio2019}
Baglio, M.~C., Russell, D.~M., Bramich, D., \& Lewis, F. 2019, ATel, 12491, 1

\bibitem[{Baglio {et~al.}(2020)Baglio, Russell, Bramich, Saikia, Pirbhoy, \&
  Lewis}]{Baglio2020}
Baglio, M.~C., Russell, D.~M., Bramich, D.~M., {et~al.} 2020, ATel, 13710

\bibitem[{Bandyopadhyay {et~al.}(1999)Bandyopadhyay, Shahbaz, Charles, \&
  Naylor}]{Bandyopadhyay1999}
Bandyopadhyay, R.~M., Shahbaz, T., Charles, P.~A., \& Naylor, T. 1999, MNRAS,
  306, 417

\bibitem[{Beals(1931)}]{Beals1931}
Beals, C. S.~. 1931, MNRAS, 91, 966

\bibitem[{Beer \& Podsiadlowski(2002)}]{Beer2002}
Beer, M.~E. \& Podsiadlowski, P. 2002, MNRAS, 331, 351

\bibitem[{Belloni {et~al.}(2011)Belloni, Motta, \&
  Mu{\~{n}}oz-Darias}]{Belloni2011}
Belloni, T.~M., Motta, S.~E., \& Mu{\~{n}}oz-Darias, T. 2011, Bull. Astron.
  Soc. India, 39, 409

\bibitem[{Belloni {et~al.}(2020)Belloni, Zhang, Kylafis, Reig, \&
  Altamirano}]{Belloni2020}
Belloni, T.~M., Zhang, L., Kylafis, N.~D., Reig, P., \& Altamirano, D. 2020,
  MNRAS, 496, 4366

\bibitem[{Buisson {et~al.}(2021)Buisson, Altamirano, {Armas Padilla},
  Arzoumanian, Bult, {Castro Segura}, Charles, Degenaar, {D{\'{I}}az Trigo},
  {Van Den Eijnden}, Fogantini, Gandhi, Gendreau, Hare, Homan, Knigge,
  Malacaria, Mendez, {Mu{\~{n}}oz Darias}, Ng, {{\"{O}}zbey Arabacl},
  Remillard, Strohmayer, Tombesi, Tomsick, Vincentelli, \&
  Walton}]{Buisson2021}
Buisson, D.~J., Altamirano, D., {Armas Padilla}, M., {et~al.} 2021, MNRAS, 503,
  5600

\bibitem[{Buisson {et~al.}(2020)Buisson, Hare, Guver, Altamirano, Gendreau,
  Arzoumanian, Bult, Strohmayer, Segura, Garcia, Remillard, Tomsick, Chenevez,
  Jaisawal, Arabaci, Vincentelli, Homan, Guillot, Wolff, Chakrabarty, \&
  Ng}]{Buisson2020b}
Buisson, D. J.~K., Hare, J., Guver, T., {et~al.} 2020, ATel, 13563, 1

\bibitem[{Buxton \& Vennes(2003)}]{Buxton2003}
Buxton, M. \& Vennes, S. 2003, MNRAS, 342, 105

\bibitem[{Carotenuto {et~al.}(2021)Carotenuto, Corbel, Tremou, Russell,
  Tzioumis, Fender, Woudt, Motta, Miller-Jones, Chauhan, Tetarenko, Sivakoff,
  Heywood, Horesh, {Van Der Horst}, Koerding, \& Mooley}]{Carotenuto2021}
Carotenuto, F., Corbel, S., Tremou, E., {et~al.} 2021, MNRAS, 504, 444

\bibitem[{Carotenuto {et~al.}(2022)Carotenuto, Tetarenko, \&
  Corbel}]{Carotenuto2022}
Carotenuto, F., Tetarenko, A.~J., \& Corbel, S. 2022, MNRAS, 000, 1

\bibitem[{Casares {et~al.}(1991)Casares, Charles, Jones, Rutten, \&
  Callanan}]{Casares1991}
Casares, J., Charles, P.~A., Jones, D. H.~P., Rutten, R. G.~M., \& Callanan,
  P.~J. 1991, MNRAS, 250, 712

\bibitem[{Casares {et~al.}(1995)Casares, Marsh, Charles, Martin, Martin,
  Harlaftis, Pavlenko, \& Wagner}]{Casares1995}
Casares, J., Marsh, T.~R., Charles, P.~A., {et~al.} 1995, MNRAS, 274, 565

\bibitem[{Casares {et~al.}(2019)Casares, Mu{\~{n}}oz-Darias, {Mata
  S{\'{a}}nchez}, Charles, Torres, {Armas Padilla}, Fender, \&
  Garc{\'{i}}a-Rojas}]{Casares2019}
Casares, J., Mu{\~{n}}oz-Darias, T., {Mata S{\'{a}}nchez}, D., {et~al.} 2019,
  MNRAS, 488, 1356

\bibitem[{{Castro Segura} {et~al.}(2022){Castro Segura}, Knigge, Long,
  Altamirano, {Armas Padilla}, Bailyn, Buckley, Buisson, Casares, Charles,
  Combi, C{\'{u}}neo, Degenaar, del Palacio, {D{\'{i}}az Trigo}, Fender,
  Gandhi, Georganti, Guti{\'{e}}rrez, {Hernandez Santisteban},
  Jim{\'{e}}nez-Ibarra, Matthews, M{\'{e}}ndez, Middleton, Mu{\~{n}}oz-Darias,
  {{\"{O}}zbey Arabacı}, Pahari, Rhodes, Russell, Scaringi, van~den Eijnden,
  Vasilopoulos, Vincentelli, \& Wiseman}]{CastroSegura2022}
{Castro Segura}, N., Knigge, C., Long, K.~S., {et~al.} 2022, Nature, 603, 52

\bibitem[{Chaty {et~al.}(2003)Chaty, Charles, Mart{\'{i}}, Mirabel,
  Rodr{\'{i}}guez, \& Shahbaz}]{Chaty2003}
Chaty, S., Charles, P.~A., Mart{\'{i}}, J., {et~al.} 2003, MNRAS, 343, 169

\bibitem[{Chauhan {et~al.}(2021)Chauhan, Miller-Jones, Raja, Allison, Jacob,
  Anderson, Carotenuto, Corbel, Fender, Hotan, Whiting, Woudt, Koribalski, \&
  Mahony}]{Chauhan2021}
Chauhan, J., Miller-Jones, J.~C., Raja, W., {et~al.} 2021, MNRAS Lett., 501, 60

\bibitem[{Chevalier \& Ilovaisky(1991)}]{Chevalier1991}
Chevalier, C. \& Ilovaisky, S.~A. 1991, A{\&}A, 251, L11

\bibitem[{Corral-Santana {et~al.}(2016)Corral-Santana, Casares, Munoz-Darias,
  Bauer, Martinez-Pais, \& Russell}]{Corral-Santana2016}
Corral-Santana, J.~M., Casares, J., Munoz-Darias, T., {et~al.} 2016, A{\&}A,
  587, A61

\bibitem[{Corral-Santana {et~al.}(2013)Corral-Santana, Casares,
  Mu{\~{n}}oz-Darias, Rodr{\'{i}}guez-Gil, Shahbaz, Torres, Zurita, \&
  Tyndall}]{CorralSantana2013}
Corral-Santana, J.~M., Casares, J., Mu{\~{n}}oz-Darias, T., {et~al.} 2013, Sci,
  339, 1048

\bibitem[{C{\'{u}}neo {et~al.}(2020{\natexlab{a}})C{\'{u}}neo, Alabarta, Zhang,
  Altamirano, M{\'{e}}ndez, {Armas Padilla}, Remillard, Homan, Steiner, Combi,
  Mu{\~{n}}oz-Darias, Gendreau, Arzoumanian, Stevens, Loewenstein, Tombesi,
  Bult, Fabian, Buisson, Neilsen, \& Basak}]{Cuneo2020}
C{\'{u}}neo, V.~A., Alabarta, K., Zhang, L., {et~al.} 2020{\natexlab{a}},
  MNRAS, 496, 1001

\bibitem[{C{\'{u}}neo {et~al.}(2020{\natexlab{b}})C{\'{u}}neo,
  Mu{\~{n}}oz-Darias, S{\'{a}}nchez-Sierras, Jim{\'{e}}nez-Ibarra, Padilla,
  Buckley, Casares, Charles, Corral-Santana, Fender, Fern{\'{a}}ndez-Ontiveros,
  S{\'{a}}nchez, Panizo-Espinar, Ponti, \& Torres}]{Cuneo2020b}
C{\'{u}}neo, V.~A., Mu{\~{n}}oz-Darias, T., S{\'{a}}nchez-Sierras, J., {et~al.}
  2020{\natexlab{b}}, MNRAS, 498, 25

\bibitem[{{D{\'{i}}az Trigo} \& Boirin(2016)}]{DiazTrigo2016}
{D{\'{i}}az Trigo}, M. \& Boirin, L. 2016, Astron. Nachrichten, 337, 368

\bibitem[{{D{\'{i}}az Trigo} {et~al.}(2012){D{\'{i}}az Trigo}, Sidoli, Boirin,
  \& Parmar}]{DiazTrigo2012}
{D{\'{i}}az Trigo}, M., Sidoli, L., Boirin, L., \& Parmar, A.~N. 2012, A{\&}A,
  543, A50

\bibitem[{Done {et~al.}(2007)Done, Gierli{\'{n}}ski, \& Kubota}]{Done2007}
Done, C., Gierli{\'{n}}ski, M., \& Kubota, A. 2007, A{\&}AR, 15, 1

\bibitem[{Fender \& Mu{\~{n}}oz-Darias(2016)}]{Fender2016}
Fender, R. \& Mu{\~{n}}oz-Darias, T. 2016, in Lect. Notes Phys., ed. F.~Haardt,
  V.~Gorini, U.~Moschella, A.~Treves, \& M.~Colpi, Vol. 905 (Cham: Springer
  International Publishing), 65

\bibitem[{Fender {et~al.}(2004)Fender, Belloni, \& Gallo}]{Fender2004}
Fender, R.~P., Belloni, T.~M., \& Gallo, E. 2004, MNRAS, 355, 1105

\bibitem[{Fomalont {et~al.}(2001)Fomalont, Geldzahler, \&
  Bradshaw}]{Fomalont2001}
Fomalont, E.~B., Geldzahler, B.~J., \& Bradshaw, C.~F. 2001, ApJ, 558, 283

\bibitem[{Gallo {et~al.}(2003)Gallo, Fender, \& Pooley}]{Gallo2003}
Gallo, E., Fender, R.~P., \& Pooley, G.~G. 2003, MNRAS, 344, 60

\bibitem[{Gilfanov(2010)}]{Gilfanov2010}
Gilfanov, M. 2010, jet Paradig. Lect. Notes Phys., 794, 17

\bibitem[{Gottlieb {et~al.}(1975)Gottlieb, Wright, \& Liller}]{Gottlieb1975}
Gottlieb, E.~W., Wright, E.~L., \& Liller, W. 1975, ApJ, 195, L33

\bibitem[{Heida {et~al.}(2017)Heida, Jonker, Torres, \& Chiavassa}]{Heida2017}
Heida, M., Jonker, P.~G., Torres, M. A.~P., \& Chiavassa, A. 2017, ApJ, 846,
  132

\bibitem[{Higginbottom {et~al.}(2019)Higginbottom, Knigge, Long, Matthews, \&
  Parkinson}]{Higginbottom2019}
Higginbottom, N., Knigge, C., Long, K.~S., Matthews, J.~H., \& Parkinson, E.~J.
  2019, MNRAS, 484, 4635

\bibitem[{Homan {et~al.}(2021)Homan, Gendreau, Sanna, Jaisawal, Buisson, Bult,
  Altamirano, \& Neilsen}]{Homan2021}
Homan, J., Gendreau, K.~C., Sanna, A., {et~al.} 2021, ATel, 14606

\bibitem[{Homan {et~al.}(2016)Homan, Neilsen, Allen, Chakrabarty, Fender,
  Fridriksson, Remillard, \& Schulz}]{Homan2016}
Homan, J., Neilsen, J., Allen, J.~L., {et~al.} 2016, ApJ Lett., 830, L5

\bibitem[{Homan {et~al.}(2001)Homan, Wijnands, van~der Klis, Belloni, van
  Paradijs, Klein‐Wolt, Fender, \& Mendez}]{Homan2001}
Homan, J., Wijnands, R., van~der Klis, M., {et~al.} 2001, ApJS, 132, 377

\bibitem[{Iaria {et~al.}(2014)Iaria, {Di Salvo}, Burderi, Riggio, D'A{\`{i}},
  \& Robba}]{Iaria2014}
Iaria, R., {Di Salvo}, T., Burderi, L., {et~al.} 2014, A{\&}A, 561, A99

\bibitem[{Jenniskens \& D{\'{e}}sert(1994)}]{Jenniskens1994}
Jenniskens, P. \& D{\'{e}}sert, F.-X. 1994, A{\&}AS, 106, 9

\bibitem[{Jim{\'{e}}nez-Ibarra {et~al.}(2019)Jim{\'{e}}nez-Ibarra,
  Mu{\~{n}}oz-Darias, Casares, Padilla, \& Corral-Santana}]{JimenezIbarra2019b}
Jim{\'{e}}nez-Ibarra, F., Mu{\~{n}}oz-Darias, T., Casares, J., Padilla, M.~A.,
  \& Corral-Santana, J.~M. 2019, MNRAS, 489, 3420

\bibitem[{Kaur {et~al.}(2012)Kaur, Kaper, Ellerbroek, Russell, Altamirano,
  Wijnands, Yang, D'Avanzo, {de Ugarte Postigo}, Flores, Fynbo, Goldoni,
  Th{\"{o}}ne, van~der Horst, van~der Klis, Kouveliotou, Wiersema, \&
  Kuulkers}]{Kaur2012}
Kaur, R., Kaper, L., Ellerbroek, L.~E., {et~al.} 2012, ApJ, 746, L23

\bibitem[{Kausch {et~al.}(2015)Kausch, Noll, Smette, Kimeswenger, Barden,
  Szyszka, Jones, Sana, Horst, \& Kerber}]{Kausch2015}
Kausch, W., Noll, S., Smette, A., {et~al.} 2015, A{\&}A, 576, A78

\bibitem[{Khargharia {et~al.}(2010)Khargharia, Froning, \&
  Robinson}]{Khargharia2010}
Khargharia, J., Froning, C.~S., \& Robinson, E.~L. 2010, ApJ, 716, 1105

\bibitem[{Lamer {et~al.}(2021)Lamer, Schwope, Predehl, Traulsen, Wilms, \&
  Freyberg}]{Lamer2021}
Lamer, G., Schwope, A.~D., Predehl, P., {et~al.} 2021, A{\&}A, 647

\bibitem[{Lindstr{\o}m {et~al.}(2005)Lindstr{\o}m, Griffin, Kiss, Uemura,
  Derekas, M{\'{e}}sz{\'{a}}ros, \& Sz{\'{e}}kely}]{Lindstrom2005}
Lindstr{\o}m, C., Griffin, J., Kiss, L.~L., {et~al.} 2005, MNRAS, 363, 882

\bibitem[{Macdonald {et~al.}(2014)Macdonald, Bailyn, Buxton, Cantrell,
  Chatterjee, Kennedy-Shaffer, Orosz, Markwardt, \& Swank}]{MacDonald2014}
Macdonald, R.~K., Bailyn, C.~D., Buxton, M., {et~al.} 2014, Astrophys. J., 784,
  2

\bibitem[{{Mata S{\'{a}}nchez} {et~al.}(2018){Mata S{\'{a}}nchez},
  Mu{\~{n}}oz-Darias, Casares, Charles, {Armas Padilla},
  Fern{\'{a}}ndez-Ontiveros, Jim{\'{e}}nez-Ibarra, Jonker, Linares, Torres,
  Shaw, Rodr{\'{i}}guez-Gil, {Van Grunsven}, Blay, Caballero-Garc{\'{i}}a,
  Castro-Tirado, Chinchilla, Farina, Ferragamo, Lopez-Martinez,
  Rubi{\~{n}}o-Martin, \& Su{\'{a}}rez-Andr{\'{e}}s}]{MataSanchez2018}
{Mata S{\'{a}}nchez}, D., Mu{\~{n}}oz-Darias, T., Casares, J., {et~al.} 2018,
  MNRAS, 481, 2646

\bibitem[{Mata-S{\'{a}}nchez {et~al.}(2015)Mata-S{\'{a}}nchez,
  Mu{\~{n}}oz-Darias, Casares, Corral-Santana, \& Shahbaz}]{MataSanchez2015}
Mata-S{\'{a}}nchez, D., Mu{\~{n}}oz-Darias, T., Casares, J., Corral-Santana,
  J.~M., \& Shahbaz, T. 2015, MNRAS, 454, 2199

\bibitem[{{Mata S{\'{a}}nchez} {et~al.}(2017){Mata S{\'{a}}nchez},
  Mu{\~{n}}oz-Darias, Casares, \& Jim{\'{e}}nez-Ibarra}]{MataSanchez2017}
{Mata S{\'{a}}nchez}, D., Mu{\~{n}}oz-Darias, T., Casares, J., \&
  Jim{\'{e}}nez-Ibarra, F. 2017, MNRAS, 464, L41

\bibitem[{{Mata S{\'{a}}nchez} {et~al.}(2015){Mata S{\'{a}}nchez},
  Mu{\~{n}}oz-Darias, Casares, Steeghs, {Ramos Almeida}, \& {Acosta
  Pulido}}]{MataSanchez2015b}
{Mata S{\'{a}}nchez}, D., Mu{\~{n}}oz-Darias, T., Casares, J., {et~al.} 2015,
  MNRAS Lett., 449, L1

\bibitem[{{Mata S{\'{a}}nchez} {et~al.}(2022){Mata S{\'{a}}nchez},
  Mu{\~{n}}oz-Darias, C{\'{u}}neo, {Armas Padilla}, S{\'{a}}nchez-Sierras,
  Panizo-Espinar, Casares, Corral-Santana, \& Torres}]{MataSanchez2022a}
{Mata S{\'{a}}nchez}, D., Mu{\~{n}}oz-Darias, T., C{\'{u}}neo, V.~A., {et~al.}
  2022, ApJL, 926, L10

\bibitem[{Matsuoka {et~al.}(2009)Matsuoka, Kawasaki, Ueno, Tomida, Kohama,
  Suzuki, Adachi, Ishikawa, Mihara, Sugizaki, Isobe, Nakagawa, Tsunemi, Miyata,
  Kawai, Kataoka, Morii, Yoshida, Negoro, Nakajima, Ueda, Chujo, Yamaoka,
  Yamazaki, Nakahira, You, Ishiwata, Miyoshi, Eguchi, Hiroi, Katayama, \&
  Ebisawa}]{Matsuoka2009}
Matsuoka, M., Kawasaki, K., Ueno, S., {et~al.} 2009, PASJ, 61, 999

\bibitem[{McClintock \& Remillard(2006)}]{McClintock2006}
McClintock, J.~E. \& Remillard, R.~A. 2006, in Compact Stellar X-ray Sources,
  Vol.~39 (Cambridge, UK: Cambridge University Press), 157--214

\bibitem[{Mu{\~{n}}oz-Darias {et~al.}(2020)Mu{\~{n}}oz-Darias, {Armas Padilla},
  Jim{\'{e}}nez-Ibarra, Panizo-Espinar, Casares, Altamirano, Buisson, Segura,
  C{\'{u}}neo, Degenaar, Fogantini, Knigge, {Mata S{\'{a}}nchez}, {Ozbey
  Arabaci}, S{\'{a}}nchez-Sierras, Torres, {Van Den Eijnden}, \&
  Vincentelli}]{Munoz-Darias2020}
Mu{\~{n}}oz-Darias, T., {Armas Padilla}, M., Jim{\'{e}}nez-Ibarra, F., {et~al.}
  2020, ApJ, 893, L19

\bibitem[{Mu{\~{n}}oz-Darias {et~al.}(2016)Mu{\~{n}}oz-Darias, Casares, {Mata
  S{\'{a}}nchez}, Fender, {Armas Padilla}, Linares, Ponti, Charles, Mooley, \&
  Rodriguez}]{Munoz-Darias2016}
Mu{\~{n}}oz-Darias, T., Casares, J., {Mata S{\'{a}}nchez}, D., {et~al.} 2016,
  Nat, 534, 75

\bibitem[{Mu{\~{n}}oz-Darias {et~al.}(2017)Mu{\~{n}}oz-Darias, Casares, {Mata
  S{\'{a}}nchez}, Fender, {Armas Padilla}, Mooley, Hardy, Charles, Ponti,
  Motta, Dhillon, Gandhi, Jim{\'{e}}nez-Ibarra, Butterley, Carey, Grainge,
  Hickish, Littlefair, Perrott, Razavi-Ghods, Rumsey, Scaife, Scott,
  Titterington, \& Wilson}]{Munoz-Darias2017}
Mu{\~{n}}oz-Darias, T., Casares, J., {Mata S{\'{a}}nchez}, D., {et~al.} 2017,
  MNRAS, 465, L124

\bibitem[{Mu{\~{n}}oz-Darias {et~al.}(2013)Mu{\~{n}}oz-Darias, Coriat, Plant,
  Ponti, Fender, \& Dunn}]{Munoz-Darias2013}
Mu{\~{n}}oz-Darias, T., Coriat, M., Plant, D.~S., {et~al.} 2013, MNRAS, 432,
  1330

\bibitem[{Mu{\~{n}}oz-Darias {et~al.}(2014)Mu{\~{n}}oz-Darias, Fender, Motta,
  \& Belloni}]{Munoz-Darias2014}
Mu{\~{n}}oz-Darias, T., Fender, R.~P., Motta, S.~E., \& Belloni, T.~M. 2014,
  MNRAS, 443, 3270

\bibitem[{Mu{\~{n}}oz-Darias {et~al.}(2019)Mu{\~{n}}oz-Darias,
  Jim{\'{e}}nez-Ibarra, Panizo-Espinar, Casares, S{\'{a}}nchez, Ponti, Fender,
  Buckley, Garnavich, Torres, Padilla, Charles, Corral-Santana, Kajava, Kotze,
  Littlefield, S{\'{a}}nchez-Sierras, Steeghs, \& Thomas}]{Munoz-Darias2019}
Mu{\~{n}}oz-Darias, T., Jim{\'{e}}nez-Ibarra, F., Panizo-Espinar, G., {et~al.}
  2019, ApJ, 879, L3

\bibitem[{Mu{\~{n}}oz-Darias {et~al.}(2018)Mu{\~{n}}oz-Darias, Torres, \&
  Garcia}]{Munoz-Darias2018}
Mu{\~{n}}oz-Darias, T., Torres, M. A.~P., \& Garcia, M.~R. 2018, MNRAS, 479,
  3987

\bibitem[{Nakahira {et~al.}(2019)Nakahira, Kawai, Negoro, Nakajima, Sakamaki,
  Maruyama, Aoki, Kobayashi, Mihara, Yatabe, Takao, Matsuoka, Sakamoto, Serino,
  Sugita, Hashimoto, Yoshida, Sugizaki, Tachibana, Morita, Ueno, Tomida,
  Ishikawa, Sugawara, Isobe, Shimomukai, Midooka, Ueda, Tanimoto, Morita,
  Yamada, Ogawa, Tsuboi, Iwakiri, Sasaki, Kawai, Sato, Tsunemi, Yoneyama,
  Asakura, Ide, Yamauchi, Hidaka, Iwahori, Kawamuro, Yamaoka, Shidatsu, \&
  Kawakubo}]{Nakahira2019}
Nakahira, S., Kawai, N., Negoro, H., {et~al.} 2019, ATel, 12469, 1

\bibitem[{Negoro {et~al.}(2020)Negoro, Nakajima, Aoki, Kobayashi, {R. Takagi},
  Asakura, U.), Mihara, Guo, Zhou, Tamagawa, Matsuoka, Sakamoto, Serino,
  Sugita, Nishida, Yoshida, Tsuboi, Iwakiri, {R. Sasaki}, \&
  Sugizaki}]{Negoro2020}
Negoro, H., Nakajima, M., Aoki, M., {et~al.} 2020, ATel, 13994

\bibitem[{Neilsen \& Lee(2009)}]{Neilsen2009}
Neilsen, J. \& Lee, J.~C. 2009, Nat, 458, 481

\bibitem[{Orosz {et~al.}(2001)Orosz, Kuulkers, van~der Klis, McClintock,
  Garcia, Callanan, Bailyn, Jain, \& Remillard}]{Orosz2001}
Orosz, J.~A., Kuulkers, E., van~der Klis, M., {et~al.} 2001, ApJ, 555, 489

\bibitem[{Panizo-Espinar {et~al.}(2021)Panizo-Espinar, Mu{\~{n}}oz-Darias,
  {Armas Padilla}, Jim{\'{e}}nez-Ibarra, Casares, {Mata S{\'{a}}nchez},
  Panizo-Espinar, Mu{\~{n}}oz-Darias, {Armas Padilla}, Jim{\'{e}}nez-Ibarra,
  Casares, \& {Mata S{\'{a}}nchez}}]{Panizo-Espinar2021}
Panizo-Espinar, G., Mu{\~{n}}oz-Darias, T., {Armas Padilla}, M., {et~al.} 2021,
  A{\&}A, 650, A135

\bibitem[{Pirbhoy {et~al.}(2020)Pirbhoy, Baglio, Russell, Bramich, Saikia, {Al
  Yazeedi}, \& Lewis}]{Pirbhoy2020}
Pirbhoy, S.~F., Baglio, M.~C., Russell, D., {et~al.} 2020, ATel, 13451

\bibitem[{Ponti {et~al.}(2012)Ponti, Fender, Begelman, Dunn, Neilsen, \&
  Coriat}]{Ponti2012}
Ponti, G., Fender, R.~P., Begelman, M.~C., {et~al.} 2012, MNRAS, 422, 11

\bibitem[{Ponti {et~al.}(2014)Ponti, Mu{\~{n}}oz-Darias, \& Fender}]{Ponti2014}
Ponti, G., Mu{\~{n}}oz-Darias, T., \& Fender, R.~P. 2014, MNRAS, 444, 1829

\bibitem[{Prinja {et~al.}(1996)Prinja, Fullerton, \& Crowther}]{Prinja1996}
Prinja, R.~K., Fullerton, A.~W., \& Crowther, P.~A. 1996, A{\&}A, 311, 264

\bibitem[{Rahoui {et~al.}(2014)Rahoui, Coriat, \& Lee}]{Rahoui2014}
Rahoui, F., Coriat, M., \& Lee, J.~C. 2014, MNRAS, 442, 1610

\bibitem[{Russell {et~al.}(2019{\natexlab{a}})Russell, Baglio, Lewis, Russell,
  Baglio, \& Lewis}]{Russell2019c}
Russell, D.~M., Baglio, C.~M., Lewis, F., {et~al.} 2019{\natexlab{a}}, ATel,
  12439, 1

\bibitem[{Russell {et~al.}(2019{\natexlab{b}})Russell, Bramich, Lewis,
  AlMannaei, {Al Qaissieh}, {Al Qasim}, {Al Yazeedi}, Baglio, Bernardini,
  Elgalad, Gabuya, Lasota, Palado, Roche, Shivkumar, Udrescu, \&
  Zhang}]{Russell2019}
Russell, D.~M., Bramich, D.~M., Lewis, F., {et~al.} 2019{\natexlab{b}}, Astron.
  Nachrichten, 340, 278

\bibitem[{Russell {et~al.}(2011)Russell, Miller-Jones, Maccarone, Yang, Fender,
  \& Lewis}]{Russell2011}
Russell, D.~M., Miller-Jones, J. C.~A., Maccarone, T.~J., {et~al.} 2011, A{\&}A
  Lett., 739, 19

\bibitem[{S{\'{a}}nchez-Sierras \&
  Mu{\~{n}}oz-Darias(2020)}]{Sanchez-Sierras2020}
S{\'{a}}nchez-Sierras, J. \& Mu{\~{n}}oz-Darias, T. 2020, A{\&}A, 640, L3

\bibitem[{Sanna {et~al.}(2019)Sanna, Uttley, Altamirano, Homan, Jaisawal,
  Gendreau, Arzoumanian, Guver, Bozzo, Ferrigno, Papitto, Burderi, Riggio, {Di
  Salvo}, Miller, Guillot, Neilsen, Sanna, Uttley, Altamirano, Homan, Jaisawal,
  Gendreau, Arzoumanian, Guver, Bozzo, Ferrigno, Papitto, Burderi, Riggio, {Di
  Salvo}, Miller, Guillot, \& Neilsen}]{Sanna2019}
Sanna, A., Uttley, P., Altamirano, D., {et~al.} 2019, ATel, 12447, 1

\bibitem[{Shakura \& Sunyaev(1973)}]{ShakuraSunyaev1973}
Shakura, N.~I. \& Sunyaev, R.~A. 1973, in X- Gamma-Ray Astron. IAU Symp.,
  Vol.~55, 155

\bibitem[{Smak(1969)}]{Smak1969}
Smak, J. 1969, AcA, 19, 155

\bibitem[{Smette {et~al.}(2015)Smette, Sana, Noll, Horst, Kausch, Kimeswenger,
  Barden, Szyszka, Jones, Gallenne, Vinther, Ballester, \& Taylor}]{Smete2015}
Smette, A., Sana, H., Noll, S., {et~al.} 2015, A{\&}A, 576, A77

\bibitem[{Soria {et~al.}(2000)Soria, Wu, \& Hunstead}]{Soria2000}
Soria, R., Wu, K., \& Hunstead, R.~W. 2000, ApJ, 539, 445

\bibitem[{Torres {et~al.}(2021)Torres, Jonker, Casares, Miller-Jones, \&
  Steeghs}]{Torres2021}
Torres, M.~A., Jonker, P.~G., Casares, J., Miller-Jones, J.~C., \& Steeghs, D.
  2021, MNRAS, 501, 2174

\bibitem[{Torres {et~al.}(2015)Torres, Jonker, Miller-Jones, Steeghs, Repetto,
  \& Wu}]{Torres2015}
Torres, M.~A., Jonker, P.~G., Miller-Jones, J.~C., {et~al.} 2015, MNRAS, 450,
  4292

\bibitem[{Torres {et~al.}(2002)Torres, Callanan, Garcia, McClintock, Garnavich,
  Balog, Berlind, Brown, Calkins, \& Mahdavi}]{Torres2002}
Torres, M. A.~P., Callanan, P.~J., Garcia, M.~R., {et~al.} 2002, ApJ, 569, 423

\bibitem[{Torres {et~al.}(2020)Torres, Casares, Jim{\'{e}}nez-Ibarra,
  {\'{A}}lvarez-Hern{\'{a}}ndez, Mu{\~{n}}oz-Darias, Padilla, Jonker, \&
  Heida}]{Torres2020}
Torres, M. A.~P., Casares, J., Jim{\'{e}}nez-Ibarra, F., {et~al.} 2020, ApJ,
  893, L37

\bibitem[{Torres {et~al.}(2019)Torres, Casares, Jim{\'{e}}nez-Ibarra,
  Mu{\~{n}}oz-Darias, Padilla, Jonker, \& Heida}]{Torres2019}
Torres, M. A.~P., Casares, J., Jim{\'{e}}nez-Ibarra, F., {et~al.} 2019, ApJ,
  882, L21

\bibitem[{Ueda {et~al.}(1998)Ueda, Inoue, Tanaka, Ebisawa, Nagase, Kotani, \&
  Gehrels}]{Ueda1998}
Ueda, Y., Inoue, H., Tanaka, Y., {et~al.} 1998, Astrophys. J., 492, 782

\bibitem[{{Vainu Bappu} \& Menzel(1954)}]{VainuBappu1954}
{Vainu Bappu}, M.~K. \& Menzel, D.~H. 1954, ApJ, 119, 508

\bibitem[{van~der Hooft {et~al.}(1998)van~der Hooft, Heemskerk, Alberts, van
  Paradijs, van~der Hooft, Heemskerk, Alberts, \& van
  Paradijs}]{VanDerHooft1998}
van~der Hooft, F., Heemskerk, M. H.~M., Alberts, F., {et~al.} 1998, A{\&}A,
  329, 538

\bibitem[{Vernet {et~al.}(2011)Vernet, Dekker, D'Odorico, Kaper, Kjaergaard,
  Hammer, Randich, Zerbi, Groot, Hjorth, Guinouard, Navarro, Adolfse, Albers,
  Amans, Andersen, Andersen, Binetruy, Bristow, Castillo, Chemla, Christensen,
  Conconi, Conzelmann, Dam, {De Caprio}, {De Ugarte Postigo}, Delabre, {Di
  Marcantonio}, Downing, Elswijk, Finger, Fischer, Flores, Fran{\c{c}}ois,
  Goldoni, Guglielmi, Haigron, Hanenburg, Hendriks, Horrobin, Horville, Jessen,
  Kerber, Kern, Kiekebusch, Kleszcz, Klougart, Kragt, Larsen, Lizon, Lucuix,
  Mainieri, Manuputy, Martayan, Mason, Mazzoleni, Michaelsen, Modigliani,
  Moehler, M{\o}ller, {Norup S{\o}rensen}, N{\o}rregaard, P{\'{e}}roux, Patat,
  Pena, Pragt, Reinero, Rigal, Riva, Roelfsema, Royer, Sacco, Santin,
  Schoenmaker, Spano, Sweers, {Ter Horst}, Tintori, Tromp, van Dael, van~der
  Vliet, Venema, Vidali, Vinther, Vola, Winters, Wistisen, Wulterkens, \&
  Zacchei}]{Vernet2011}
Vernet, J., Dekker, H., D'Odorico, S., {et~al.} 2011, A{\&}A, 536, A105

\bibitem[{Williams \& Mason(2010)}]{Williams2010}
Williams, R. \& Mason, E. 2010, Ap{\&}SS, 327, 207

\bibitem[{Xu {et~al.}(2021)Xu, Harrison, Xu, \& Harrison}]{Xu2021}
Xu, Y., Harrison, F., Xu, Y., \& Harrison, F. 2021, ATel, 14609, 1

\bibitem[{Yao {et~al.}(2021)Yao, Kulkarni, Burdge, Caiazzo, De, Dong, Fremling,
  Kasliwal, Kupfer, van Roestel, Sollerman, Bagdasaryan, Bellm, Cenko, Drake,
  Duev, Graham, Kaye, Masci, Miranda, Prince, Riddle, Rusholme, \&
  Soumagnac}]{Yao2021}
Yao, Y., Kulkarni, S.~R., Burdge, K.~B., {et~al.} 2021, ApJ, 920, 120

\bibitem[{Yatabe {et~al.}(2019)Yatabe, Negoro, Nakajima, Sakamaki, Maruyama,
  Aoki, Kobayashi, Mihara, Nakahira, Takao, Matsuoka, Sakamoto, Serino, Sugita,
  Hashimoto, Yoshida, Kawai, Sugizaki, Tachibana, Morita, Ueno, Tomida,
  Ishikawa, Sugawara, Isobe, Shimomukai, Midooka, Ueda, Tanimoto, Morita,
  Yamada, Ogawa, Tsuboi, Iwakiri, Sasaki, Kawai, Sato, Tsunemi, Yoneyama,
  Asakura, Ide, Yamauchi, Hidaka, Iwahori, Kawamuro, Yamaoka, Shidatsu,
  Kawakubo, Yatabe, Negoro, Nakajima, Sakamaki, Maruyama, Aoki, Kobayashi,
  Mihara, Nakahira, Takao, Matsuoka, Sakamoto, Serino, Sugita, Hashimoto,
  Yoshida, Kawai, Sugizaki, Tachibana, Morita, Ueno, Tomida, Ishikawa,
  Sugawara, Isobe, Shimomukai, Midooka, Ueda, Tanimoto, Morita, Yamada, Ogawa,
  Tsuboi, Iwakiri, Sasaki, Kawai, Sato, Tsunemi, Yoneyama, Asakura, Ide,
  Yamauchi, Hidaka, Iwahori, Kawamuro, Yamaoka, Shidatsu, \&
  Kawakubo}]{Yatabe2019}
Yatabe, F., Negoro, H., Nakajima, M., {et~al.} 2019, ATel, 12425, 1

\bibitem[{Zhang {et~al.}(2020)Zhang, Altamirano, C{\'{u}}neo, Alabarta, Enoto,
  Homan, Remillard, Uttley, Vincentelli, Arzoumanian, Bult, Gendreau,
  Markwardt, Sanna, Strohmayer, Steiner, Basak, \& Neilsen}]{Zhang2020}
Zhang, L., Altamirano, D., C{\'{u}}neo, V.~A., {et~al.} 2020, MNRAS, 499, 851

\bibitem[{Zhang {et~al.}(2021)Zhang, Altamirano, Uttley, Garc{\'{i}}a,
  M{\'{e}}ndez, Homan, Steiner, Alabarta, Buisson, Remillard, Gendreau,
  Arzoumanian, Markwardt, Strohmayer, Neilsen, \& Basak}]{Zhang2021}
Zhang, L., Altamirano, D., Uttley, P., {et~al.} 2021, MNRAS, 505, 3823

\bibitem[{Zhang {et~al.}(2022)Zhang, Tao, Soria, Qu, Zhang, Weng, Zhang, Wang,
  Huang, Ma, Zhang, Ge, Song, Ma, Bu, Cai, Cao, Chang, Chen, Chen, Chen, Chen,
  Chen, Cui, Du, Gao, Gao, Gu, Guan, Guo, Han, Huo, Jia, Jiang, Jin, Kong, Li,
  Li, Li, Li, Li, Li, Li, Li, Li, Liang, Liao, Liu, Liu, Liu, Liu, Liu, Lu, Lu,
  Luo, Luo, Meng, Nang, Nie, Ou, Ren, Sai, Song, Sun, Tan, Tuo, Wang, Wang,
  Wang, Wang, Wang, Wen, Wu, Wu, Wu, Xiao, Xiao, Xiong, Chen, Yang, Yang, Yang,
  Yang, Yi, Yin, Yuan, Zhang, Zhang, Zhang, Zhang, Zhang, Zhang, Zhao, Zhao,
  Zheng, Zheng, \& Zhou}]{Zhang2022}
Zhang, W., Tao, L., Soria, R., {et~al.} 2022, ApJ, 927, 210

\end{thebibliography}

\end{document}